\documentclass[aps,twocolumn,groupedaddress,floatfix,preprintnumbers,nofootinbib,eqsecnum]{revtex4}
\usepackage{amssymb,amsmath,graphicx,multirow}

\begin{document}

\title{Strategies for Probing Non-Minimal Dark Sectors at Colliders:\\ 
    The Interplay Between Cuts and Kinematic Distributions}
\author{Keith R. Dienes$^{1,2}$\footnote{E-mail address:  {\tt dienes@email.arizona.edu}},
      Shufang Su$^{1}$\footnote{E-mail address:  {\tt shufang@email.arizona.edu}},  
      Brooks Thomas$^{3}$\footnote{E-mail address:  {\tt bthomas@physics.carleton.ca}}}
\affiliation{
     $^1$ Department of Physics, University of Arizona, Tucson, AZ  85721  USA\\
     $^2$ Department of Physics, University of Maryland, College Park, MD  20742  USA\\
     $^3$ Department of Physics, Carleton University, Ottawa, ON K1S 5B6 Canada}

\begin{abstract}
In this paper, we examine the strategies and prospects for distinguishing
between traditional dark-matter models and models with non-minimal
dark sectors --- including models of Dynamical Dark Matter (DDM) --- at
hadron colliders.  For concreteness, we focus on events with two hadronic
jets and large missing transverse energy at the Large Hadron Collider (LHC).
As we discuss, simple ``bump-hunting'' searches are not sufficient;
probing non-minimal dark sectors typically requires an analysis of the
actual shapes of the distributions of relevant kinematic variables.
We therefore begin by identifying those kinematic variables
whose distributions are particularly suited to this task.  However, as
we demonstrate, this then leads to a number of additional subtleties, since
cuts imposed on the data for the purpose of background reduction can at the
same time have the unintended consequence of distorting these distributions in
unexpected ways, thereby obscuring signals of new physics.  We therefore proceed
to study the {\it correlations}\/ between several of the most popular relevant
kinematic variables currently on the market, and investigate how imposing cuts on
one or more of these variables can impact the distributions of others.
Finally, we combine our results in order to assess the prospects for distinguishing
non-minimal dark sectors in this channel at the upgraded LHC.
\end{abstract}

\maketitle

\newcommand{\newc}{\newcommand}
\newc{\gsim}{\lower.7ex\hbox{$\;\stackrel{\textstyle>}{\sim}\;$}}
\newc{\lsim}{\lower.7ex\hbox{$\;\stackrel{\textstyle<}{\sim}\;$}}
\makeatletter
\newcommand{\biggg}{\bBigg@{3}}
\newcommand{\Biggg}{\bBigg@{4}}
\makeatother

\def\vac#1{{\bf \{{#1}\}}}

\def\beq{\begin{equation}}
\def\eeq{\end{equation}}
\def\beqn{\begin{eqnarray}}
\def\eeqn{\end{eqnarray}}
\def\calM{{\cal M}}
\def\calV{{\cal V}}
\def\calF{{\cal F}}
\def\half{{\textstyle{1\over 2}}}
\def\quarter{{\textstyle{1\over 4}}}
\def\ie{{\it i.e.}\/}
\def\eg{{\it e.g.}\/}
\def\etc{{\it etc}.\/}


\def\inbar{\,\vrule height1.5ex width.4pt depth0pt}
\def\IR{\relax{\rm I\kern-.18em R}}
 \font\cmss=cmss10 \font\cmsss=cmss10 at 7pt
\def\IQ{\relax{\rm I\kern-.18em Q}}
\def\IZ{\relax\ifmmode\mathchoice
 {\hbox{\cmss Z\kern-.4em Z}}{\hbox{\cmss Z\kern-.4em Z}}
 {\lower.9pt\hbox{\cmsss Z\kern-.4em Z}}
 {\lower1.2pt\hbox{\cmsss Z\kern-.4em Z}}\else{\cmss Z\kern-.4em Z}\fi}
\def\TBBN{T_{\mathrm{BBN}}}
\def\OmegaCDM{\Omega_{\mathrm{CDM}}}
\def\OmegaDM{\Omega_{\mathrm{CDM}}}
\def\Omegatot{\Omega_{\mathrm{tot}}}
\def\rhocrit{\rho_{\mathrm{crit}}}
\def\tnow{t_{\mathrm{now}}}
\def\arcsinh{\mbox{arcsinh}}
\def\Omegatotnow{\Omega_{\mathrm{tot}}^\ast}
\def\mij{m_{jj}}
\def\mijmin{m_{jj}^{(\mathrm{min})}}
\def\mijmax{m_{jj}^{(\mathrm{max})}}
\def\mmax{m_{\mathrm{max}}}
\def\epsig{\epsilon_{\mathrm{sig}}}
\def\Lint{\mathcal{L}_{\mathrm{int}}}
\def\MT2{M_{T2}}
\def\MTtwomax{M_{T2}^{\mathrm{max}}}
\def\BR{\mathrm{BR}}
\newcommand{\Dsle}[1]{\hskip 0.09 cm \slash\hskip -0.23 cm #1}
\newcommand{\Dirsl}[1]{\hskip 0.09 cm \slash\hskip -0.20 cm #1}
\newcommand{\met}{{\Dsle E_T}}


\input epsf





\section{Introduction\label{sec:Intro}}


Overwhelming evidence now suggests~\cite{reviews} that non-baryonic 
dark matter contributes a substantial fraction $\OmegaDM \approx 0.26$~\cite{Planck} 
of the energy density in the universe.  Experimental and observational 
data significantly constrain the fundamental properties of the particle(s) 
which contribute toward this dark-matter abundance.  Nevertheless,
a broad range of viable theoretical possibilities exist for what the dark matter 
in our universe might be.  One possibility is that the dark sector is ``minimal'' in
the sense that a single, stable particle species contributes essentially the entirety 
of $\OmegaDM$.  However, there also exist additional well-motivated possibilities 
in which the dark sector manifests a richer and more complicated {\it non-minimal}\/ 
structure.  For example, several particle species could contribute non-trivially 
toward $\OmegaDM$~\cite{MultiComponentBlock,ProfumoUbaldi,DDMBlock}.  
Indeed, $\OmegaDM$ could even represent the collective contribution 
from a vast ensemble of potentially unstable individual particle species
whose lifetimes are balanced against their cosmological abundances  --- a possibility 
known as Dynamical Dark Matter (DDM)~\cite{DDMBlock}.  Other 
extensions of the minimal case exist as well.  Thus, once an unambiguous signal of 
dark matter is identified, differentiating between all of these possibilities 
will become the next crucial task for dark-matter phenomenology.

This task presents a unique set of challenges.  Practically by definition, 
the dark sector comprises neutral particle species with similar or identical 
quantum numbers under the Standard-Model (SM) gauge group.  Indeed, many theoretical
realizations of the dark sector differ from one another only in the multiplicity and/or 
masses of such particle species and the strengths of their couplings to other fields 
in the theory.  For this reason, evidence for a particular structure within the dark sector
is not usually expected to manifest itself via the simultaneous observation of signal
excesses in multiple detection channels.  Rather, such evidence will often appear
only in the shapes of the distributions of particular kinematic quantities 
in one particular channel.  Such distributions include, for example, the 
recoil-energy spectra obtained from direct-detection experiments, the energy spectra 
of photons or other cosmic-ray particles at indirect-detection experiments, and the 
distributions of a number of kinematic variables (particle momenta, invariant and 
transverse masses, \etc) at colliders.  Of course, some information about the properties 
of the dark particles can be ascertained merely by identifying the kinematic endpoints 
of these distributions.  However, such information is typically insufficient to distinguish 
single-particle from multi-particle dark sectors.  Indeed, for such purposes, an analysis 
of the full {\it shape}\/ of the distribution is required.  

In many experimental contexts, the extraction of information from kinematic distributions is 
complicated by the presence of sizable backgrounds --- backgrounds which can only be reduced 
through the imposition of stringent event-selection criteria.  While such cuts are often 
critical for signal extraction, they can have unintended consequences for distribution-based
searches.  Specifically, the cuts imposed on one variable can potentially distort the shapes 
of the kinematic distributions of other variables whenever those variables are non-trivially 
correlated.  Such effects are not particularly important in ``bump-hunting'' searches, 
in which the goal is merely to identify an excess in the total number of observed events 
over the expected background.  By contrast, in distribution-based searches, these effects 
can obscure critical information and lead to misleading results --- or, in certain cases,
can actually amplify distinctive features which point toward different kinds of dark-sector 
non-minimality.  These issues are especially relevant for collider searches, wherein a variety of 
different strategies often exist for extracting signal from background in any particular 
channel.  Indeed, different strategies may offer very different prospects for 
distinguishing among different dark-matter scenarios. 

Effective strategies for distinguishing non-minimality in the dark sector
been developed for certain detection channels.  For example, it has been shown that 
DDM ensembles can give rise to statistically significant deviations from the kinematic 
distributions associated with traditional, single-particle dark-matter candidates at 
colliders~\cite{DDMLHC1}, at direct-detection experiments~\cite{DDDDM1}, and at 
cosmic-ray detectors~\cite{DDMAMS}.  Similar analyses have also been performed for 
other non-minimal dark-matter scenarios in the context of direct 
detection~\cite{ProfumoUbaldi}, indirect detection~\cite{MiXDM},
and collider searches~\cite{KaustubhZ3Cusps,KaustubhZ3MT2,MT2VariantsAsymmetricDecays}.
However, in general, it is also important to investigate the effects of correlations 
between kinematic variables and their impact on distribution shapes.

In this paper, we investigate the prospects for distinguishing between minimal and 
non-minimal dark sectors on the basis of kinematic distributions at the the CERN Large 
Hadron Collider (LHC).  For concreteness, we focus on the dijet~+~$\met$ channel, primarily 
due to its kinematic simplicity and its relevance for a wide variety of new-physics scenarios, 
including supersymmetry, theories with universal extra dimensions, and theories including 
scalar leptoquarks.  Nevertheless, we emphasize that many of our findings transcend this 
particular analysis and apply more broadly to any distribution-based search for dark-sector 
non-minimality.  

This paper is organized as follows.  In Sect.~\ref{sec:Preliminaries},
we introduce the class of minimal and non-minimal dark-matter models which
will serve as benchmarks in our study.  In Sect.~\ref{sec:VarsDDMvsTDM}, we then
review the properties of various kinematic variables which can be constructed 
for the dijet~+~$\met$ channel.  We examine the kinematic distributions of 
these variables for both our minimal and non-minimal benchmark models and
determine the degree to which each variable is sensitive to the structure of
the dark sector.  In Sect.~\ref{sec:CorrelBetweenVars}, we proceed to examine the 
correlations between these different variables and provide a qualitative assessment 
of how cuts on certain variables affect the distributions of other variables.       
In Sect.~\ref{sec:results}, we combine our results in order to assess 
the extent to which signal-event distributions can be used to differentiate DDM ensembles 
from traditional dark-matter candidates.  Finally, in Sect.~\ref{sec:conclusions}, we 
conclude with a discussion of how to extend our analysis of cuts and correlations 
among collider variables to other channels relevant for the detection and 
differentiation of dark-matter candidates.

Before proceeding, one final comment is in order.
Our primary aim in this paper is to examine the information that different
kinematic variables can provide about the structure of the dark sector 
and to assess the impact of correlations between these variables. In particular,
it is not our aim to present an exhaustive quantitative analysis of the 
discovery prospects for dark-sector non-minimality in the dijet~+~$\met$ channel.
For this reason, we choose to focus on the signal contributions to the event 
rate and treat the SM backgrounds merely as a motivation for the cuts we impose 
on the signal distributions.  However, we note that substantial residual backgrounds 
from processes such as $\bar{t}t$~+~jets, $W^\pm$~+~jets, and $Z$~+~jets remain for
this channel even after stringent cuts are applied.  These residual backgrounds make 
extracting information about the dark sector particularly challenging.  
In Sect.~\ref{sec:conclusions}, we shall return to this issue and discuss potential 
techniques for further reducing these backgrounds in future 
collider searches for non-minimal dark sectors.  These issues notwithstanding, 
we emphasize that the correlations we discuss here are every bit as relevant 
for a full study including
both signal and background contributions as they are for this background-free 
analysis.  Moreover, many of the general considerations we discuss here transcend 
this particular channel and apply more broadly to any search which involves the 
analysis of kinematic distributions rather than merely the identification 
of an excess in the number of observed events.


\section{Preliminaries: Parametrizing the Dark Sector \label{sec:Preliminaries}}


As discussed in the Introduction, our goal is to examine the strategies and prospects for 
distinguishing non-minimal dark sectors on the basis of 
results in the dijet~+~$\met$ channel at the LHC.~  
However, dark sectors can exhibit various degrees of non-minimality
ranging from just a few dark particles all the way to large DDM-like ensembles.
In order to obtain a sense of the full scope of possibilities, in this paper we shall therefore
consider two extremes which sit at opposite poles of complexity.

The simplest situation one can consider is the case of a single dark-matter particle 
$\chi$ of mass $m_\chi$.  For concreteness, we take $\chi$ to be a Dirac fermion 
which transforms as a singlet under the SM gauge group.  We also assume that the theory 
contains an additional scalar field $\phi$ with mass $m_\phi > m_\chi$ which transforms 
in the fundamental representation of $SU(3)_c$ and which can therefore be produced 
copiously via strong interactions at the LHC.~  In addition, we also assume that $\phi$ 
couples to $\chi$ and right-handed SM quarks $q_R$ via an interaction Lagrangian of the 
form 
\begin{equation}
  \mathcal{L}_{\mathrm{int}} ~=~ \sum_q \big[ c_{\chi q} \phi^\dagger\bar{\chi} q_R 
    + \mathrm{h.c.}\big]~,
  \label{eq:LintTrad}
\end{equation}  
where the $c_{\chi q}$ are dimensionless coupling coefficients.  
For simplicity, we take the $c_{\chi q}$ to be real and focus on the case in which $\phi$ 
couples to a single light quark species, which we here take to 
be the up quark --- \ie, we take $c_{\chi u} \equiv c_\chi$ to be non-vanishing, 
while $c_{\chi q} = 0$ for $q \in \{d,s,c,b,t\}$.  Such coupling structures arise 
naturally for a variety of exotic particles in well-motivated extensions of the 
SM, including up squarks in flavor-aligned supersymmetry.  Note that in general
the $c_{\chi q}$ may be non-vanishing for either up-type or down-type quarks,
depending on the $U(1)_{\mathrm{EM}}$ charge of $\phi$, but
not for both simultaneously.  Furthermore, we assume 
that there are no other interactions within this simplified model which contribute to 
$\phi$ decay, so that the decay process $\phi\rightarrow q\overline{\chi}$ dominates the width 
$\Gamma_\phi$ of $\phi$.  This situation arises naturally in any scenario in which there 
exists a symmetry under which $\phi$ and $\chi$ transform non-trivially but all SM 
particles transform trivially, thereby rendering $\chi$ stable.  
Finally, we assume that the characteristic timescale $\tau_\phi$ associated with this 
decay process is sufficiently short ($\tau_\phi \lesssim 10^{-12}$~s) that $\phi$ 
decays promptly within a collider detector once it is produced.    

At the opposite extreme of non-minimality, we shall consider a benchmark scenario in which the 
dark sector consists of an entire ensemble of individual components --- \eg, an entire so-called
``DDM ensemble''~\cite{DDMBlock}.  Indeed, in this paper such DDM ensembles will be taken
as our canonical representatives of highly non-minimal dark sectors.  The class of DDM models 
on which we choose to focus is that in which the constituent particles $\chi_n$ of this 
ensemble, $n=1,\ldots,N$, are SM-gauge-singlet Dirac fermions with a mass spectrum of the form 
\begin{equation}
  m_n ~=~ m_0 + n^\delta \Delta m~,
  \label{eq:MassSpectrum}
\end{equation}
where the mass $m_0$ of the lightest constituent in the ensemble, the mass-splitting parameter 
$\Delta m$, and the power-law index $\delta$ are free parameters of the theory.  Note that
in this parametrization, $\Delta m > 0$ and $\delta > 0$ by construction, so that the index
$n$ labels the $\chi_n$ in order of increasing mass.  Indeed, it turns out that many 
naturally-occurring DDM ensembles have mass spectra of this form.  

Just as for the single-component case discussed above, we also assume that the theory
includes an additional, heavy scalar field $\phi$ with mass $m_\phi$ which transforms in the 
fundamental representation of the SM $SU(3)_c$ gauge group.  Likewise, the $\chi_n$ are assumed 
to couple to this heavy scalar and to right-handed quarks $q_R$ via the interaction Lagrangian 
\begin{equation}
  \mathcal{L}_{\mathrm{int}} ~=~ \sum_{n=0}^N \sum_q \big[  c_{n q} \phi^\dagger
     \bar{\chi}_n q_R  + \mathrm{h.c.}\big]~.
  \label{eq:Lint}
\end{equation}   
Indeed, this is the analogue of the interaction Lagrangian in Eq.~(\ref{eq:LintTrad}) 
for the single-particle case, and the $c_{n q}$ are a set of dimensionless coupling 
coefficients analogous to $c_\chi$.  Once again, to facilitate direct comparison 
with the single-particle case, we focus on the case in which the $c_{n q}$ are real 
and in which only $c_{n u} \equiv c_n$ is non-vanishing for each $n$, while 
$c_{n q} = 0$ for all other quark species.  We also likewise assume that decay processes
of the form $\phi\rightarrow q\overline{\chi}_n$ dominate $\Gamma_\phi$ to such an extent 
that all other contributions to that width can safely be neglected, and that all such 
decay processes occur promptly within the detector.  Finally, we assume that the $c_n$ 
scale across the ensemble according to a power-law relation of the form 
\begin{equation}
  c_n ~=~ c_0 \left(\frac{m_n}{m_0}\right)^\gamma~,
\end{equation}
where the exponent $\gamma$ is another free parameter of the theory.
     
In general, the number of dark-matter components in the ensemble can be quite large, 
and it is possible for the masses of the heaviest components in the ensemble to greatly 
exceed $m_\phi$.  However, only those states $\chi_n$ with $m_n < m_\phi$ can be produced 
through the decays of $\phi$, and thus only those states will be relevant for the study 
in this paper.  As a result, for the purposes of this study, we shall effectively consider 
$N$ to be the number of states in the ensemble with masses less than $m_\phi$, with the 
understanding that $m_N < m_\phi$ but $m_{N+1}\geq m_\phi$.      
     
We emphasize that we have chosen to focus on a scenario in which $\phi$ is a 
Lorentz scalar and transforms in the fundamental representation of $SU(3)_c$ merely 
for concreteness.  Similar results will emerge in any alternative 
scenario in which $\phi$ transforms under the Lorentz and $SU(3)_c$ groups in
such a way that two-body decays to a SM quark or gluon and one of the $\chi_n$ 
are permitted and dominate $\Gamma_\phi$.  For example, the results obtained for 
an $SU(3)_c$-octet fermion whose width is dominated by decays of the form 
$\phi \rightarrow g \chi_n$, where $g$ denotes a SM gluon, are quite similar to 
those we obtain in this paper.  Moreover, the enhanced pair-production cross-section 
in this case would improve the prospects for distinguishing dark-sector non-minimality.
Similar results would likewise emerge for different assignments of the 
coupling coefficients $c_{n q}$.


\section{Probing the Dark Sector:  
  Variables that Do, and Variables that Don't\label{sec:VarsDDMvsTDM}}


In this section, we survey some of the kinematic variables in common use
for the event topology discussed in Sect.~\ref{sec:Preliminaries}, 
with an eye towards assessing their possible utility 
in probing non-minimal dark sectors.  As discussed in the Introduction, bump-hunting is 
not enough --- we need to analyze the shapes of the {\it distributions}\/ associated with 
these variables.  In this connection, several questions emerge.  For instance, is it 
possible to distinguish non-minimal or multi-component dark sectors from 
traditional single-component dark sectors on the basis of such distributions?  
If so, which kinematic variables provide the best prospects for doing this? 
In short, we seek to understand the extent to which differences in the particle content 
of the dark sector can affect the shapes of the kinematic distributions of whatever 
useful variables can be constructed.

In this section, we take the first step toward answering these questions by examining
the qualitative features associated with the kinematic distributions of different 
variables and assessing to what extent these features are affected by dark-sector
non-minimality.  We begin by enumerating the kinematic variables we consider in this 
study and discussing their general properties.  We then discuss the underlying methodology 
and assumptions inherent in our calculation of the corresponding kinematic distributions 
using Monte-Carlo techniques.  Finally, we present the results of this calculation 
and provide a preliminary assessment as to which variables have distributions  
which are sensitive to non-minimality, and which do not.

\subsection{Kinematic Variables}
 
As discussed in the Introduction, one of the challenges of extracting information
about the dark sector from the dijet~+~$\met$ channel is the paucity of information 
contained in the description of any given event.  Indeed,
other than variables that characterize the angular size and 
substructure of the two jets (considerations which are not particularly relevant for this 
analysis), such events are completely characterized by only six independent degrees of freedom: 
the six components of the momenta $\vec{p}_1$ and $\vec{p}_2$ of the jets $j_1$ and $j_2$,
respectively.  Nevertheless, a number of kinematic variables can be constructed from these 
six degrees of freedom which can be used to extract information from this channel.  These 
include
\begin{itemize}
  \item The magnitude $\met$ of the missing transverse momentum in the event.
  \item The magnitudes $p_{T_1}$ and $p_{T_2}$ of the transverse momenta of 
    the leading jet $j_1$ and next-to-leading jet $j_2$ in the event, respectively, 
    where the jets are ranked by $p_T$.
  \item The scalar sum $H_{T_{jj}}$ of the transverse momenta $p_{T_1}$ and $p_{T_2}$.
  \item The scalar sum $H_T$ of $\met$ and the transverse momenta $p_{T_1}$ and $p_{T_2}$.  
  \item The absolute value $|\Delta\phi_{jj}|$ of the angle between $j_1$ and $j_2$.
  \item The variable $\alpha_{T} \equiv p_{T_2}/m_{jj}$, where 
    $m_{jj}$ is the invariant mass of $j_1$ and $j_2$.  This
    variable was introduced in Ref.~\cite{RandallDijetVariables} and is correlated with 
    the degree to which these two leading jets are back-to-back.
  \item The transverse mass $M_{T_1}$ constructed from $\vec{p}_{T_1}$ and 
    the total missing-transverse-momentum vector ${\Dirsl\vec{p}_T}$. 
  \item The standard $\MT2$ variable~\cite{MT2}.  
\end{itemize}

The last variable, $\MT2$, will play a significant role in this paper.  We therefore 
pause to discuss its definition and properties in some detail.  This quantity is essentially
a generalization of the transverse-mass variable for use in situations in which more
than one invisible particle is present in the final state for a given collider process.
For the process $pp\rightarrow \phi^\dagger\phi \rightarrow j j \chi_a \overline{\chi}_b$,
which is the primary focus of this study, this variable is defined as
\begin{equation}
  \MT2^2(\widetilde{m}) ~\equiv~ 
    \raisebox{-0.15cm}{$\stackrel{\displaystyle\mathrm{min}}
    {\scriptstyle  \not\vec{p}_{T_a} + \not\vec{p}_{T_b} = \not\vec{p}_T}$}
    \bigg[\mathrm{max}\Big\{\left(M_T^2\right)_{1a},\left(M_T^2\right)_{2b}\Big\}\bigg]~,
   \label{eq:MT2Def}
\end{equation}
where $\widetilde{m}$ is a common ``trial mass'' which is assumed for {\it both} $\chi_a$
and $\chi_b$; where $\vec{p}_{T_1}$ and $\vec{p}_{T_2}$ are the transverse momenta
of the two leading lets (ranked by $p_T$);
where $\Dirsl\vec{p}_T$ is the total missing-transverse-momentum vector for the event;
where $\Dirsl\vec{p}_{T_a}$ and $\Dirsl\vec{p}_{T_b}$ represent possible
partitions of this total missing-transverse-momentum vector between the two invisible
particles $\chi_a$ and $\chi_b$; and where
\begin{align}
  \left(M_T^2\right)_{1a} &\equiv~ 
    m_{j_1}^2 + \widetilde{m}^2
    + 2\big(E_{T_{j_1}}\Dsle E_{T_a}\! - 
    \vec{p}_{T_1}\! \cdot \Dirsl\vec{p}_{T_a}\big)
    \nonumber\\
  \left(M_T^2\right)_{2b} &\equiv~ 
    m_{j_2}^2 + \widetilde{m}^2
    + 2\big(E_{T_{j_2}}\Dsle E_{T_b}\! - 
    \vec{p}_{T_2}\! \cdot \Dirsl\vec{p}_{T_b}\big)~
  \label{eq:TransverseMasses}
\end{align}
are the squared transverse masses of $j_1$ with $\chi_a$ and of $j_2$ with $\chi_b$ for
any particular such partition, respectively.  In this expression,
$m_{j_1}$ and $m_{j_2}$ denotes the masses of the two jets (which are negligible
in practice).  Note that by construction, the transverse energies
$\Dsle E_{T_a} \equiv (|{\Dirsl\vec{p}_{T_a}}|^2 + \widetilde{m}^2)^{1/2}$
and $\Dsle E_{T_b} \equiv (|{\Dirsl\vec{p}_{T_b}}|^2 + \widetilde{m}^2)^{1/2}$
appearing in Eq.~(\ref{eq:TransverseMasses}) are both defined in terms of the trial
mass $\widetilde{m}$.  The $\MT2$ variable in Eq.~(\ref{eq:MT2Def}) is defined to be the
minimum of the greater of these two transverse masses over all possible partitions of
$\Dirsl\vec{p}_T$ between $\Dirsl\vec{p}_{T_a}$ and $\Dirsl\vec{p}_{T_b}$.

In cases in which each of the two decay chains in the event includes only one
visible-sector particle in the final state, it can be
shown~\cite{MT2,MT2PropertiesBarr,MT2PropertiesChoi} that the partition of
$\Dirsl\vec{p}_T$ for which this minimum occurs is always the so-called ``balanced''
solution --- \ie,  the solution for which
$\left(M_T^2\right)_{1a} = \left(M_T^2\right)_{2b}$.  For this balanced
solution, one finds that
\begin{align}
  \MT2^2(\widetilde{m}) &=~
    \widetilde{m}^2 + A + \big(A^2 - m_{j_1}^2 m_{j_2}^2\big)^{1/2}
    \nonumber \\ &~~~~\, \times 
    \bigg(1+\frac{4\widetilde{m}^2}{2A - m_{j_1}^2 - m_{j_2}^2}\bigg)^{1/2}~,
\end{align}
where
\begin{equation}
    A ~\equiv~ E_{T_1}E_{T_2} + \vec{p}_{T_1}\!\cdot\vec{p}_{T_2}~.
  \label{eq:DefOfA}
\end{equation}

One particularly useful feature of the $\MT2$ variable is that it is bounded from
above.  Indeed, the maximum possible value $\MT2$ can attain within any sample of events
is the one in which all of the particles involved are maximally transverse, and the
transverse mass reconstructed for each of the two decay chains in the event
coincides with its corresponding invariant mass.  It therefore follows that
in traditional dark-matter models, for which $m_a = m_b = m_\chi$ by assumption,
the maximum possible value $\MTtwomax$ for $\MT2$ is equal to the mass $m_\phi$ of
the parent particle.  However, since the value of $m_\chi$ is in general not
{\it a priori}\/ known, one can only examine the functional dependence of
$\MTtwomax$ on $\widetilde{m}$.

In DDM scenarios, in which the masses $m_a$ and $m_b$ of the invisible particles
associated with the two decay chains in a given event are not necessarily equal,
the $\MT2$ values obtained for a population of signal events in the
$p p \rightarrow jj + \met$ channel may differ significantly from those obtained
in traditional dark-matter models, even for the same $m_\phi$.
Moreover, we emphasize that the kinematic endpoint $\MTtwomax$ itself
is not particularly useful in discriminating between DDM ensembles and traditional
dark-matter candidates; rather, it is only by comparing the {\it shapes}\/ of the full
$\MT2$ distributions that one might hope to distinguish DDM ensembles from traditional
dark-matter candidates.  Of course, the maximum value of $\MT2$ for a DDM
ensemble is obtained for $m_a = m_b = m_0$, with all final-state particles
maximally transverse.  Thus, for a sufficiently large sample of events, one finds that
$\MTtwomax(m_0) \rightarrow m_\phi$, a result identical to that obtained for a
traditional dark-matter candidate with $m_\chi = m_0$.

We emphasize that the list of kinematic variables we have presented in this section
is by no means complete.  Indeed, a number of additional
collider variables have been developed to extract information from channels 
involving substantial $\met$.  These include ratios of the transverse energies of 
visible particles~\cite{TransverseEnergyRatios}, the so-called ``constransverse mass'' and
variants thereof~\cite{InvariantMassVariants}, and numerous generalizations of the 
$\MT2$ variable~\cite{MT2Variants}, including particular 
variants~\cite{MT2VariantsAsymmetricDecays} specifically designed for probing 
scenarios in which the invisible particles have unequal masses.  While we do not 
consider any of these additional variables in this paper, we note that an analysis of 
the extent to which their kinematic distributions may be influenced by different 
sets of event-selection criteria would be completely analogous to the procedure outlined 
here.

\subsection{Calculating Distributions\label{sec:CalculatingDistributions}}

Our ultimate goal is to examine the prospects these kinematic variables proffer for 
distinguishing between different dark sectors at the LHC.~  As discussed in the 
Introduction, an analysis of the full {\it distributions}\/ associated with these 
kinematic variables is frequently required for this purpose.  Thus, using
Monte-Carlo simulations, we explicitly derive kinematic distributions 
for these variables for both the traditional dark-matter candidates and the DDM 
ensembles included in this study.

Specifically, all data sets used in this study were generated at the
parton level using MadGraph~5/MadEvent~1.4.8~\cite{MadGraph5} with model files 
obtained from the FeynRules~\cite{FeynRules} package as inputs.  In order to account 
for the effects of detector uncertainties, we have smeared the original 
values for the magnitude $p_T$  of the transverse momentum, the azimuthal angle $\phi$, 
and the pseudorapidity $\eta$ obtained for each jet in each parton-level data set according 
to the following procedure.
We replace the value of each of these three jet parameters with a 
pseudorandom value distributed according to a Gaussian probability-distribution function.
We take the mean value for this Gaussian function to be the original value obtained 
from the Monte-Carlo simulation and the variance to be the square of the uncertainty 
in the measurement of the corresponding variable.  In particular, we take the uncertainty 
in the $p_T$ of each 
jet to be given by the jet-$p_T$ resolution for the CMS detector.  This resolution was 
evaluated in Ref.~\cite{CMSjetpTRes} as a function of $p_T$ and is well approximated by 
the expression 
\begin{equation}
  \delta p_T (p_T) ~ \approx ~ 0.037 
    + 0.67 \times \left(\frac{p_T}{\mathrm{GeV}}\right)^{-1/2}~.
\end{equation}
We likewise take the uncertainties in $\eta$ and $\phi$ to be given by the 
pseudorapidity and azimuthal-angle resolutions of the CMS detector, respectively.  
These resolutions were evaluated as functions of $p_T$ in Ref.~\cite{CMSjetEtaPhiRes}.  
We find that the results are well approximated by the expressions 
\begin{align}
  \delta \eta(p_T) & \approx~ 0.024 
    + 3.00 \times \left(\frac{p_T}{\mathrm{GeV}}\right)^{-3/2} \nonumber \\ &~~~~
    + 0.070 \times \left(\frac{p_T}{\mathrm{GeV}}\right)^{-1/2} \nonumber \\ 
  \delta \phi(p_T) & \approx~ 
    0.027 + 2.45 \times \left(\frac{p_T}{\mathrm{GeV}}\right)^{-1} \nonumber \\ &~~~~
    - 0.046 \times \left(\frac{p_T}{\mathrm{GeV}}\right)^{-1/2}
\end{align}
for $p_T \lesssim 1$~TeV.

In addition to the above smearing, we also incorporate a set of precuts into
our analysis.  In particular, we consider in our analysis only those events in each 
of these data sets which satisfy the following precuts, which are designed to mimic 
a realistic detector acceptance: 
\begin{itemize}
\item A transverse momentum $p_{T_{j}} \geq 40$~GeV and pseudorapidity 
  $|\eta_{j}| \leq 3$ for each of the two highest-$p_T$ jets in the event.  
\item A minimum separation $\Delta R_{jj} \geq 0.4$ between those two
  leading jets, where  
  $\Delta R_{jj} \equiv \sqrt{(\Delta \eta_{jj})^2 + (\Delta \phi_{jj})^2}$.
\end{itemize}
Note that these acceptance cuts alone do not guarantee that a particular event 
satisfies any particular detector trigger.  Indeed, we have chosen not to incorporate 
any particular triggering criteria into our precuts because different 
event-selection strategies may in principle be constructed around different triggers, 
and we wish to inject as little prejudice as possible about the event-selection strategy 
at this stage in the analysis.  However, when we turn to assess the prospects for 
distinguishing DDM ensembles from traditional dark-matter candidates in 
Sect.~\ref{sec:results}, we shall include additional cuts designed to satisfy triggering
requirements.  

\subsection{Kinematic Distributions}

Following the above procedures, we can now evaluate the distributions associated
with each of our kinematic variables.  

\begin{figure*}
\begin{center}
  \epsfxsize 2.10 truein \epsfbox {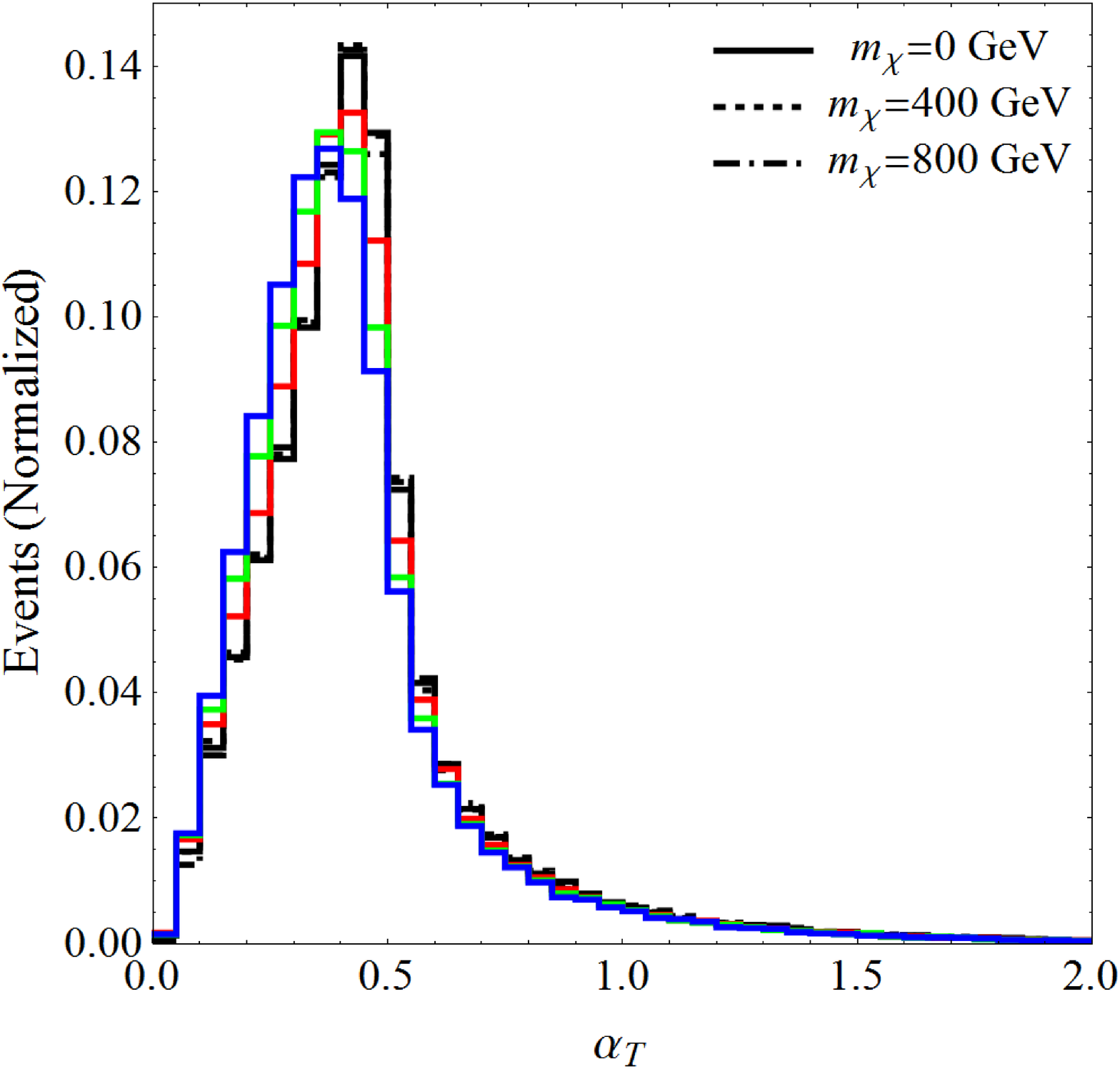}~~
  \epsfxsize 2.10 truein \epsfbox {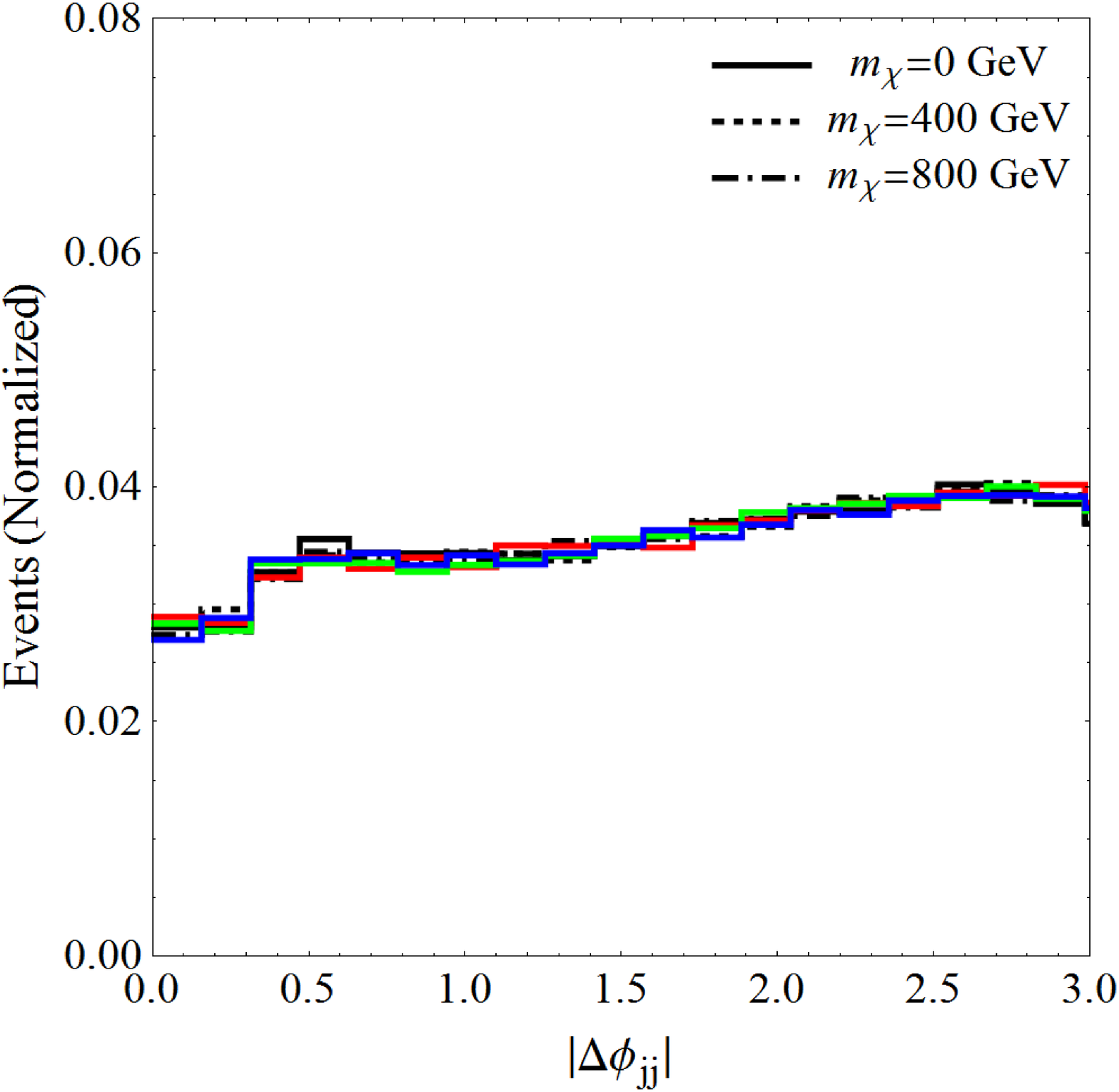}~~
  \epsfxsize 2.10 truein \epsfbox {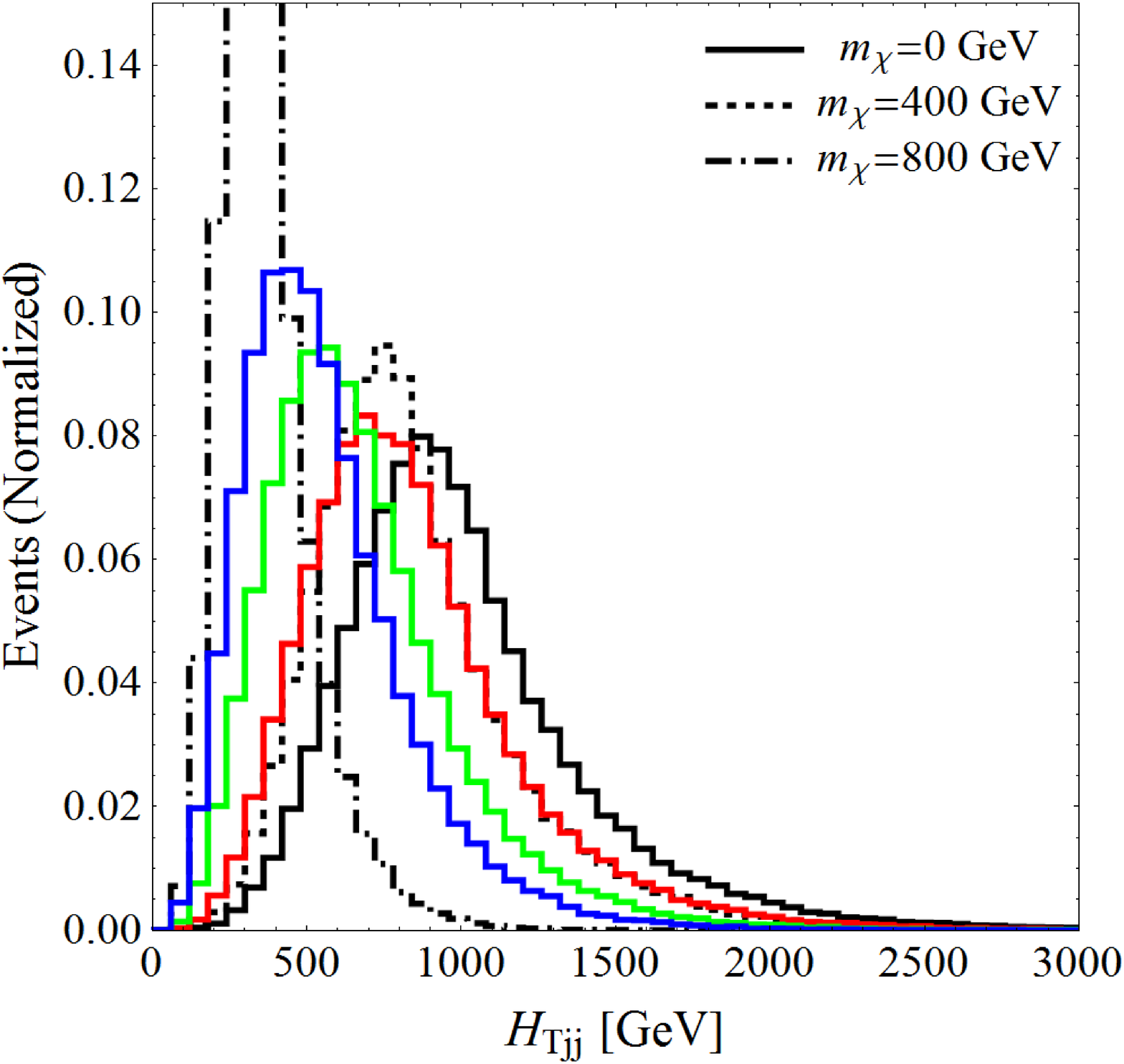}
\end{center}
\caption{ ~A comparison of the normalized $\alpha_T$ (left panel), $|\Delta \phi_{jj}|$
(center panel), and $H_{T_{jj}}$ (right panel) distributions associated 
with traditional dark-matter candidates as well as DDM ensembles.
In each panel, the black curves correspond to
distributions for a representative set of traditional dark-matter candidates,
while
the red, green, and blue curves in each panel correspond to DDM ensembles with
$m_0 = 200$~GeV, $\Delta m = 50$~GeV, $\delta = 1$, and
$\gamma = \{0,1,2\}$, respectively.   
All distributions shown correspond to a parent-particle mass $m_\phi=1$~TeV.}
\label{fig:alphaTDeltaPhiHTjjDistsForDDM}
\begin{center}
  \epsfxsize 2.10 truein \epsfbox {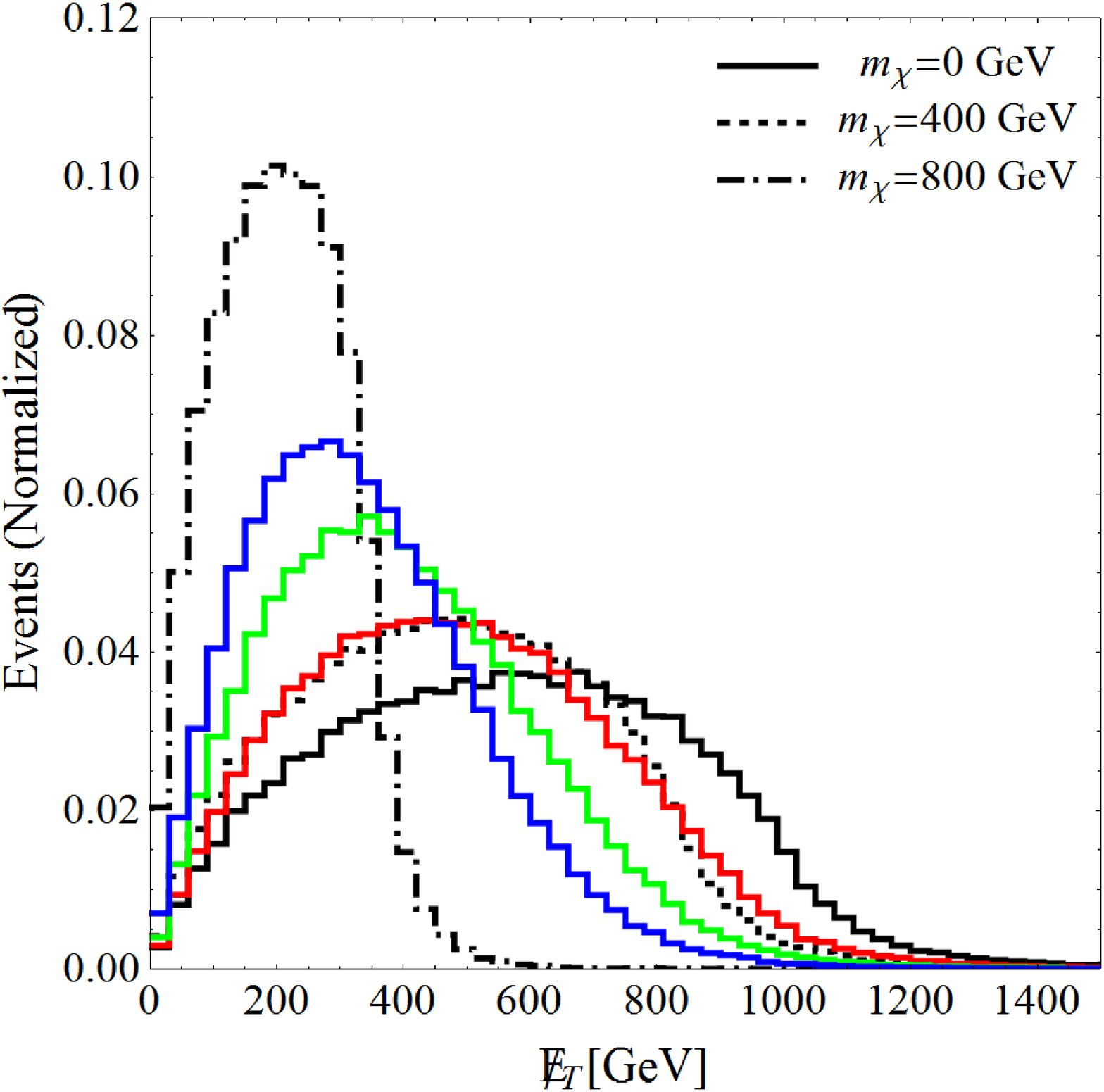}~~
  \epsfxsize 2.10 truein \epsfbox {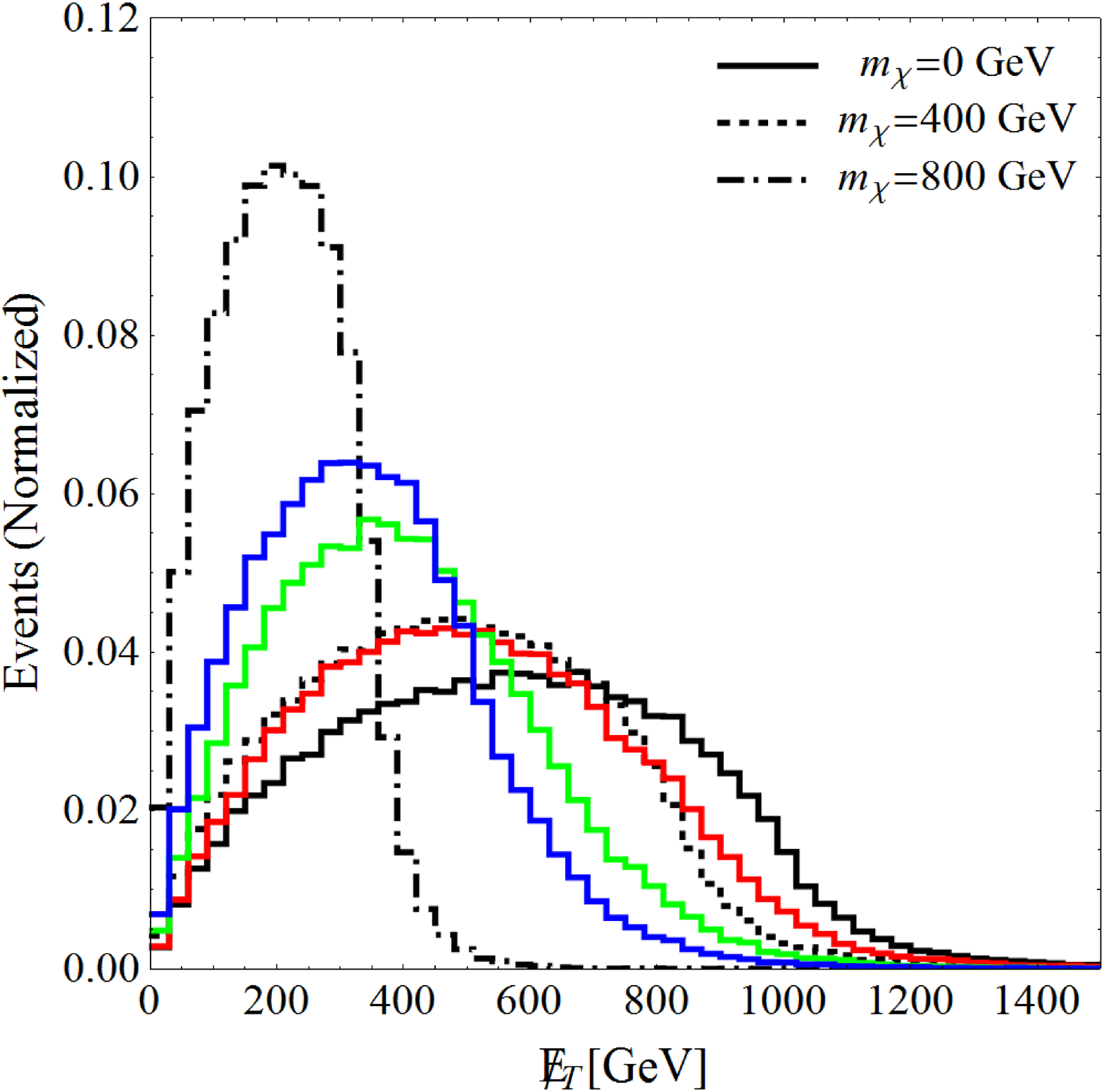}~~
  \epsfxsize 2.10 truein \epsfbox {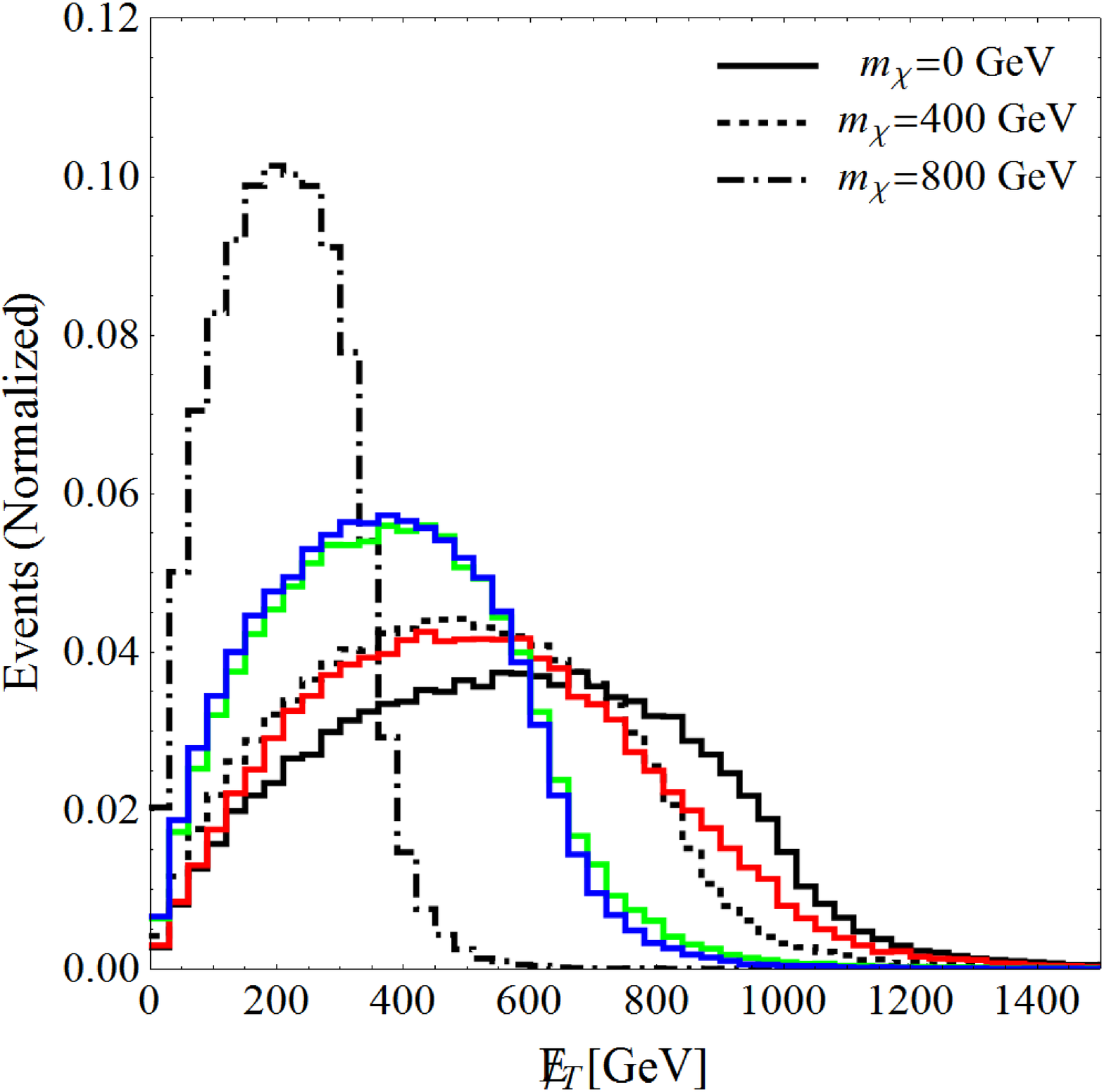}
\end{center}
\caption{ ~A comparison of the normalized $\met$ distributions associated with 
traditional dark-matter candidates as well as DDM ensembles.
In each panel, the black curves correspond to
distributions for a representative set of traditional dark-matter candidates,
while the colored curves
in the left, middle, and right panels
correspond to the DDM parameter choices 
$m_\phi=1$~TeV, $m_0 = 200$~GeV, $\delta = 1$, 
and $\Delta m = \{50,300,500\}$~GeV, respectively,
with  $\gamma=0$ (red), $\gamma=1$ (yellow), and $\gamma=2$ (blue).}
\label{fig:METDistsForDDM}
\begin{center}
  \epsfxsize 2.10 truein \epsfbox {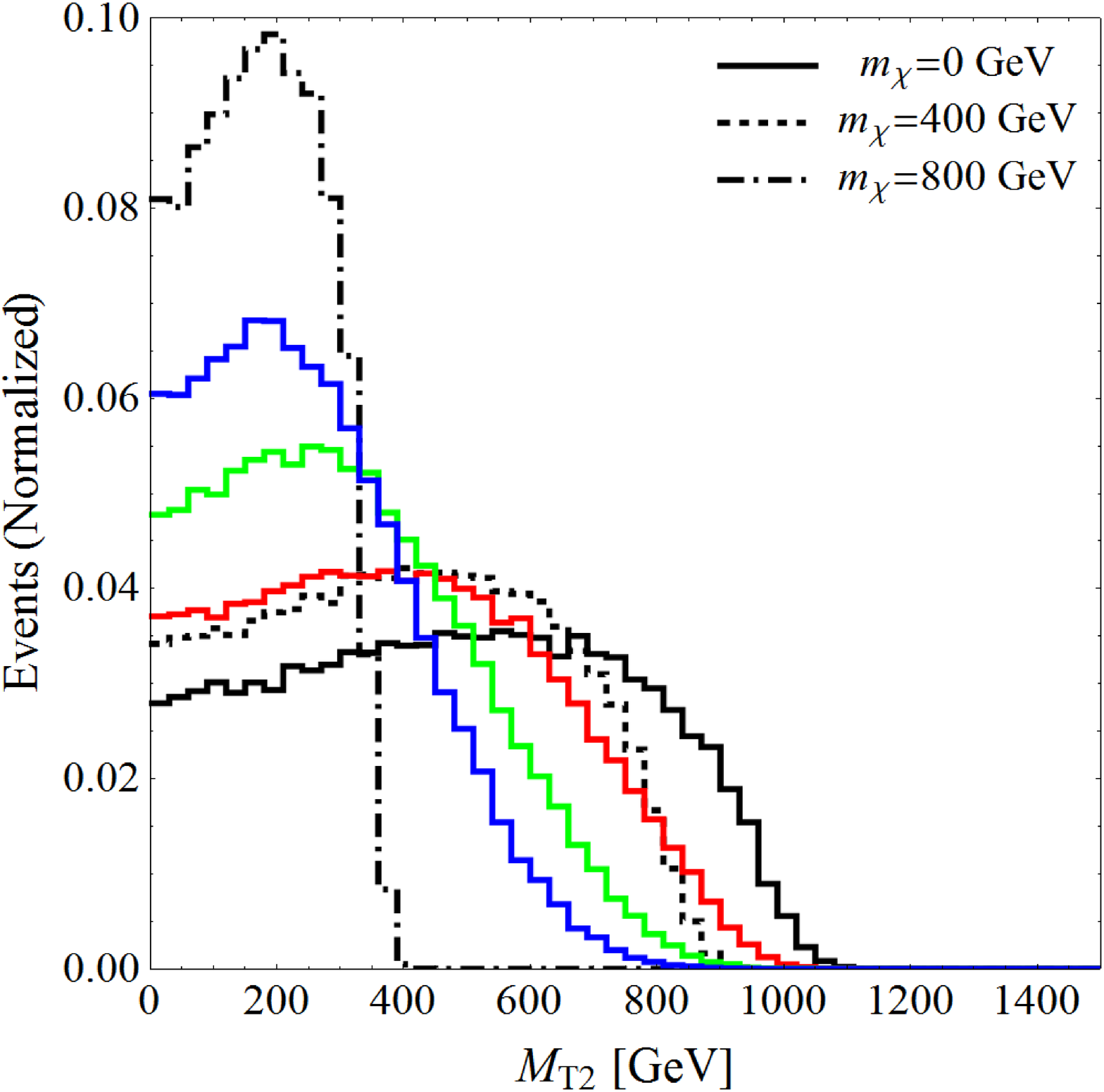}~~
  \epsfxsize 2.10 truein \epsfbox {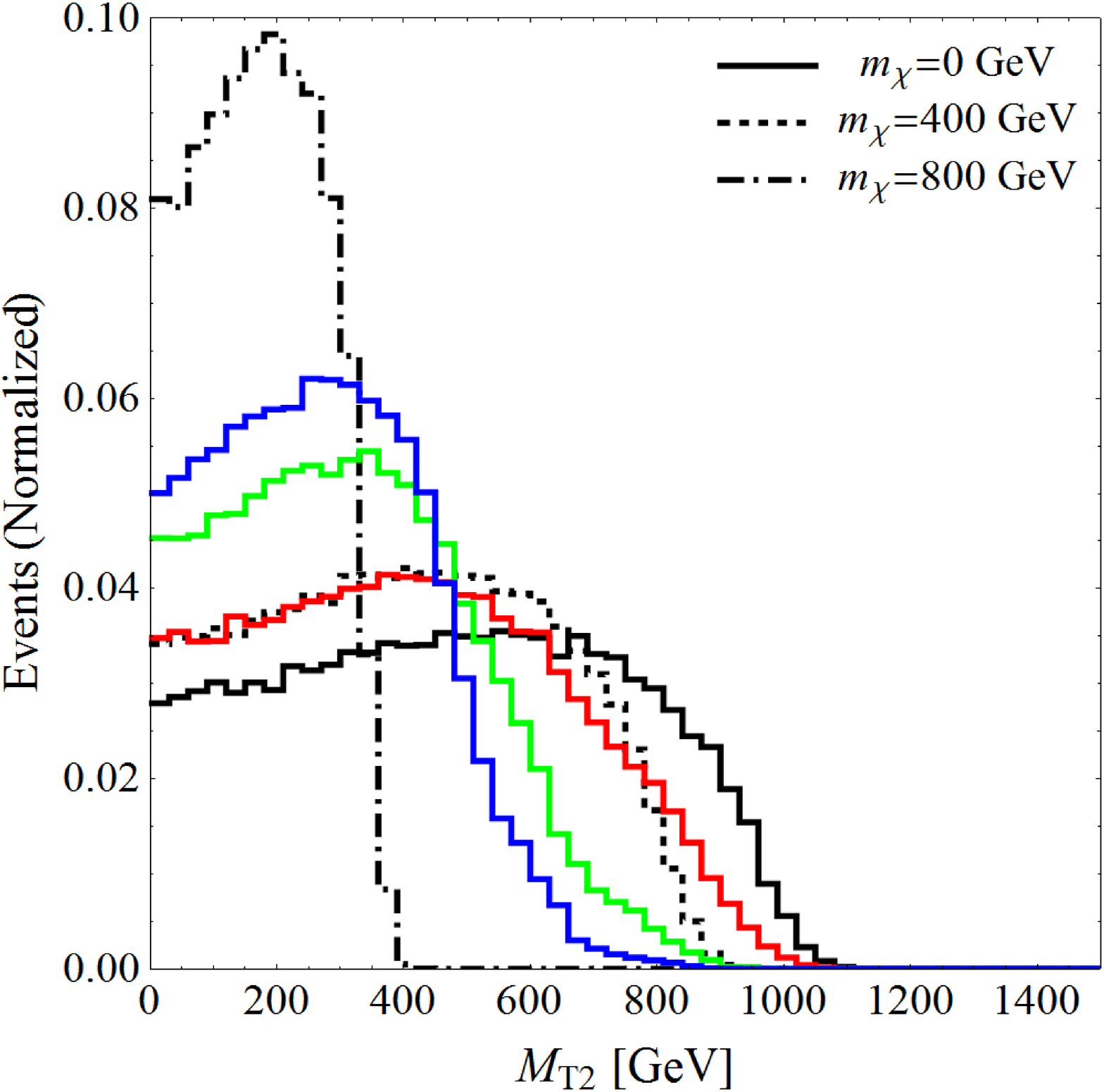}~~
  \epsfxsize 2.10 truein \epsfbox {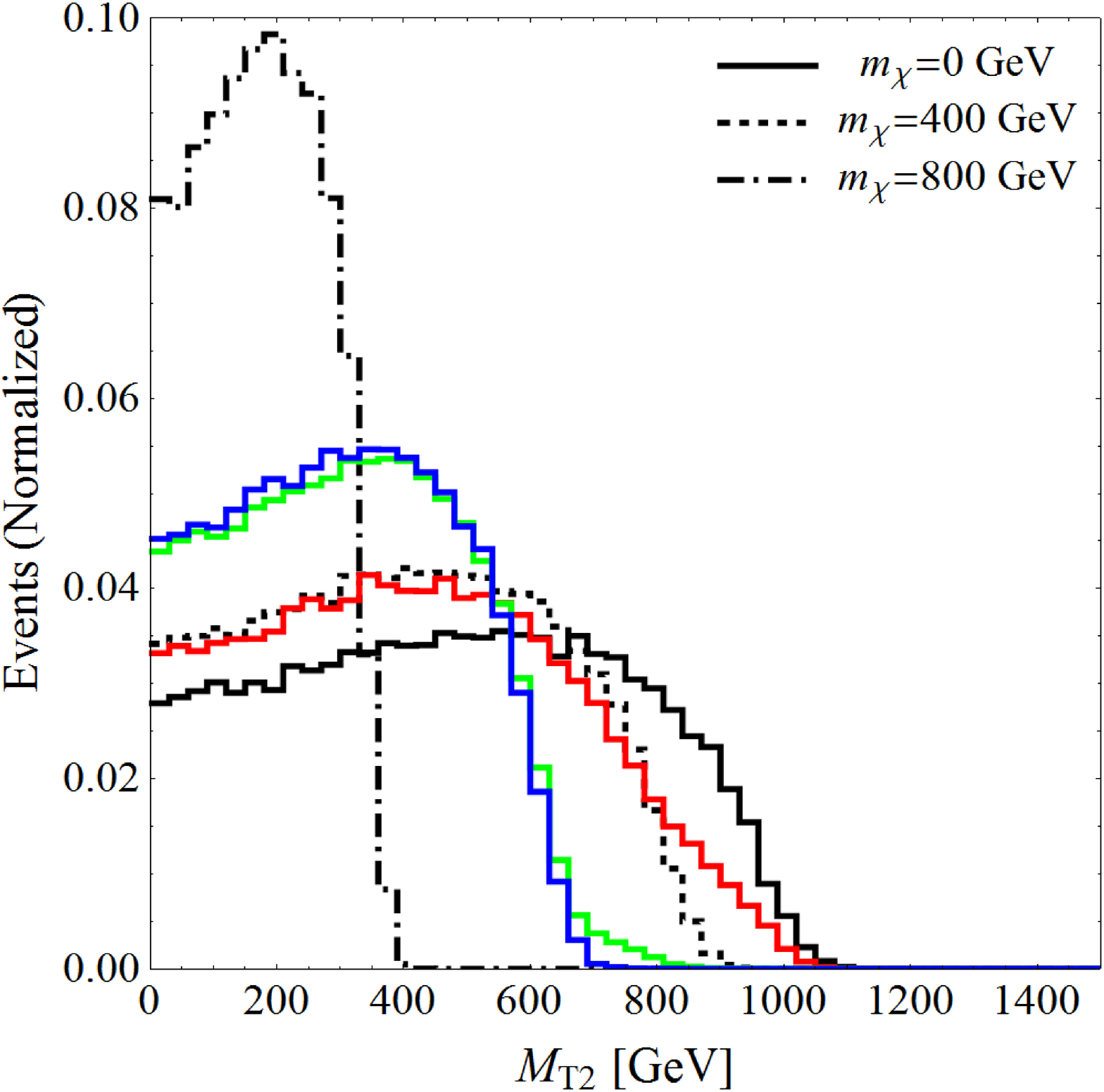}
\end{center}
\caption{ ~Same as Fig.~\ref{fig:METDistsForDDM}, but for normalized $\MT2$ distributions with trial 
  mass $\tilde m=0$.}
\label{fig:MT2DistsForDDM}
\end{figure*}

We begin by considering the distributions 
associated with the variables $\alpha_T$, $|\Delta \phi_{jj}|$, and $H_{T_{jj}}$.
Indeed, these variables are of particular relevance for new-physics searches in
the dijet~+~$\met$ channel because cuts on these variables are particularly effective 
in reducing SM backgrounds from QCD processes, electroweak processes, {\it etc}\/.
In the left panel of Fig.~\ref{fig:alphaTDeltaPhiHTjjDistsForDDM}, we
display the $\alpha_T$ distributions associated with a number of traditional 
dark-matter models characterized by different values of $m_\chi$, as well as the 
distributions associated with a number of DDM ensembles characterized by different 
values of $\gamma$ for fixed $m_0 = 100$~GeV, $\Delta m = 50$~GeV, and $\delta = 1$.
Note that these distributions
have been normalized so that the total area under each is unity.  
In the center panel of this figure, we display the $|\Delta \phi_{jj}|$
distributions corresponding to the same parameter choices.
The results shown in these two panels suggest that the shapes of $\alpha_T$ and
$|\Delta \phi_{jj}|$ distributions are not particularly sensitive to the
structure of the dark sector and therefore not particularly useful for distinguishing
among different dark-matter scenarios.  Indeed, we find that the $\alpha_T$ and
$|\Delta\phi_{jj}|$ distributions obtained for different choices of the DDM model
parameters $m_0$, $\Delta m$, and $\delta$ do not differ significantly from the
distributions shown in Fig.~\ref{fig:alphaTDeltaPhiHTjjDistsForDDM}. 

In the right panel of Fig.~\ref{fig:alphaTDeltaPhiHTjjDistsForDDM}, we display the 
$H_{T_{jj}}$ distributions associated with the same set of traditional dark-matter 
models and DDM ensembles as in the left and center panels.  In contrast to the
corresponding $\alpha_T$ and $|\Delta\phi_{jj}|$ distributions, which are largely
insensitive to the structure of the dark sector, the $H_{T_{jj}}$ distributions shown in
Fig.~\ref{fig:alphaTDeltaPhiHTjjDistsForDDM} display a somewhat greater sensitivity
to the spectrum of masses and couplings of the invisible particles.  However,
despite this sensitivity, we also find that for any given DDM ensemble,
there is generally a traditional dark-matter candidate with some value of $m_\chi$
which yields a fairly similar $H_{T_{jj}}$ distribution.

We now turn to the distributions associated with the kinematic variables 
$\met$ and $\MT2$.  In Fig.~\ref{fig:METDistsForDDM}, we display the normalized $\met$ 
distributions associated with a number of traditional dark-matter models characterized 
by different values of $m_\chi$ as well as a number of DDM ensembles characterized by 
different values of the parameters $m_0$, $\Delta m$, and $\gamma$ (with $\delta = 1$).
In contrast to the distributions for $\alpha_T$ and $|\Delta \phi_{jj}|$ displayed in
Fig.~\ref{fig:alphaTDeltaPhiHTjjDistsForDDM}, the $\met$ distributions shown in
Fig.~\ref{fig:METDistsForDDM} are far more sensitive to the structure of the dark sector.
Events which involve the heavier $\chi_n$ in a DDM ensemble tend to have smaller $\met$
values.  The contribution from such events therefore tends to shift the peak of the
distribution to lower $\met$ --- especially when $\gamma$ is large and the branching fraction
of $\phi$ to the heavier kinematically-accessible $\chi_n$ in the ensemble is sizable.
On the other hand, the contribution from the lighter $\chi_n$ nevertheless contributes to
the ``tail'' of the distribution at high $\met$.  The interplay between these two effects
results in distributions for DDM ensembles with shifted peaks and longer
tails --- distributions whose distinctive shapes are not reproduced by any traditional
dark-matter candidate, regardless of the value of $m_\chi$.  
Moreover, it can also be seen from Fig.~\ref{fig:METDistsForDDM} that the shape of the 
$\met$ distribution associated with a DDM ensemble is also sensitive to the choice of DDM 
model parameters.  Indeed, as discussed above, larger values of $\gamma$ serve to shift the 
peak of the distribution to lower $\met$.  Furthermore, comparing results across the three 
panels shown in the figure, we also see that larger values of $\Delta m$ result in a sharp
decrease in event count with increasing $\met$ above the value for which the distribution
peaks, whereas smaller values of $\Delta m$ result in a more gradual decline in event
count with increasing $\met$.

Finally, in Fig.~\ref{fig:MT2DistsForDDM}, we display the $\MT2$ distributions associated 
with a variety of DDM ensembles for the trial mass $\widetilde{m} = 0$.  Note that
this choice of $\widetilde{m}$ generally yields the most distinctive distributions
because the $\MT2$ variable is restricted to the range  
$\widetilde{m} \leq \MT2(\widetilde{m}) \leq m_\phi$ for any particular choice of 
$\widetilde{m}$.  Thus, as $\widetilde{m}$ increases, the window of possible $\MT2$ 
values narrows and the resulting distributions become 
more ``compressed'' and therefore less distinctive.

Once again, just as with the $\met$ distributions shown in
Fig.~\ref{fig:METDistsForDDM}, we find that the $\MT2$ distributions shown in 
this figure display a significant sensitivity to the structure of the dark
sector.  Moreover, the shapes of the distributions of these two variables depend 
on the DDM model parameters in similar ways.  For example, the shapes of the 
$\MT2$ distributions associated with DDM ensembles with larger values of $\gamma$
peak at lower values of $\MT2$ while still retaining a significant tail which 
extends out to the kinematic endpoint at $\MTtwomax = m_\phi$.
For large $\Delta m$, the individual contributions to the distributions from 
events with different values of $m_a$ and $m_b$ can be independently resolved, as shown in
the right panel of the figure, and a ``kink'' behavior arises similar to that which 
arises in the case in which multiple invisible particles are produced from a single 
decay chain~\cite{KaustubhZ3MT2}.  By contrast, for small $\Delta m$, these contributions
cannot be resolved, and the tail of the resulting distribution appears smooth.  In either
case, it is evident from Fig.~\ref{fig:MT2DistsForDDM} that $\MT2$ distributions 
are particularly useful for distinguishing DDM ensembles from traditional dark-matter 
candidates --- and from each other.  In fact, as we shall show in 
Sect.~\ref{sec:results}, $\MT2$ is an even better variable than $\met$ for 
extracting information about the structure of the dark sector.

The distributions associated with the other kinematic variables discussed above 
(including $p_{T_1}$, $p_{T_2}$, $H_T$, and the transverse mass $M_{T_1}$) likewise 
display some sensitivity to the structure of the dark sector.  However, we find that these 
distributions have far less power for distinguishing minimal from non-minimal
dark sectors than those associated with $\met$ and $\MT2$.  Moreover, these variables 
also turn out to be less effective than $\alpha_T$, $|\Delta \phi_{jj}|$,
and $H_{T_{jj}}$ for extracting signal from background.  Thus, we shall not
consider the distributions of these other variables further.    

In summary, we conclude that the distributions of some kinematic variables, such as $\alpha_T$ 
and $|\Delta \phi_{jj}|$, are almost completely insensitive to the degree of non-minimality 
of the dark sector.  By contrast, we find that others, such as $\met$ and $\MT2$, are 
particularly sensitive to such non-minimality.  Finally, we find that still others, such 
as $H_{T_{jj}}$, lie between these two extremes.  
 

\section{Correlations Between Kinematic Variables \label{sec:CorrelBetweenVars}}


In any experiment, signals come with unwanted backgrounds.  Finding cuts that reduce 
these backgrounds relative to the resulting signal is therefore an important task.
Although we are not performing a detailed analysis of the backgrounds in this paper,
there are certain SM backgrounds which are endemic to dark-matter searches in this
channel.  Along with these are certain cuts which are well known to be particularly 
advantageous in dealing with these backgrounds.

For example, cuts on variables such as $\alpha_T$ and $|\Delta\phi_{jj}|$  --- variables 
which are strongly correlated with the angle between the spatial momenta $\vec{p}_{T_1}$ and 
$\vec{p}_{T_2}$ of the two leading jets in a given event --- are particularly effective 
in reducing the substantial QCD background in the dijet~+~$\met$ channel.  This is 
because QCD-background events tend to be back-to-back and therefore seldom have 
$\alpha_T \gtrsim 0.5$.  Indeed, the minimum cut $\alpha_T > 0.55$ imposed in
CMS searches in this channel~\cite{CMSJetsPlusMET,CMSJetsPlusMETUpdate} has been shown 
to be extremely effective in reducing --- and indeed effectively eliminating --- the 
sizable background from QCD processes.

Likewise, cuts on variables such as $H_{T_{jj}}$ and $H_T$ --- variables which are correlated with 
the overall energy of the underlying event --- can be effective in reducing the remaining 
SM backgrounds which are dominated by processes such as $\bar{t}t$~+~jets, $W^\pm$~+~jets 
with the $W^\pm$ decaying leptonically, and $Z$~+~jets with the $Z$ decaying into a neutrino 
pair.  These variables are relevant for searches in the dijet~+~$\met$ channel for 
another reason as well: detector triggers useful in selecting events 
involving hadronic jets and substantial $\met$ frequently rely on the scalar sum of the 
transverse momenta of those jets exceeding a particular threshold.  

\begin{figure*}
\begin{center}
  \epsfxsize 2.15 truein \epsfbox {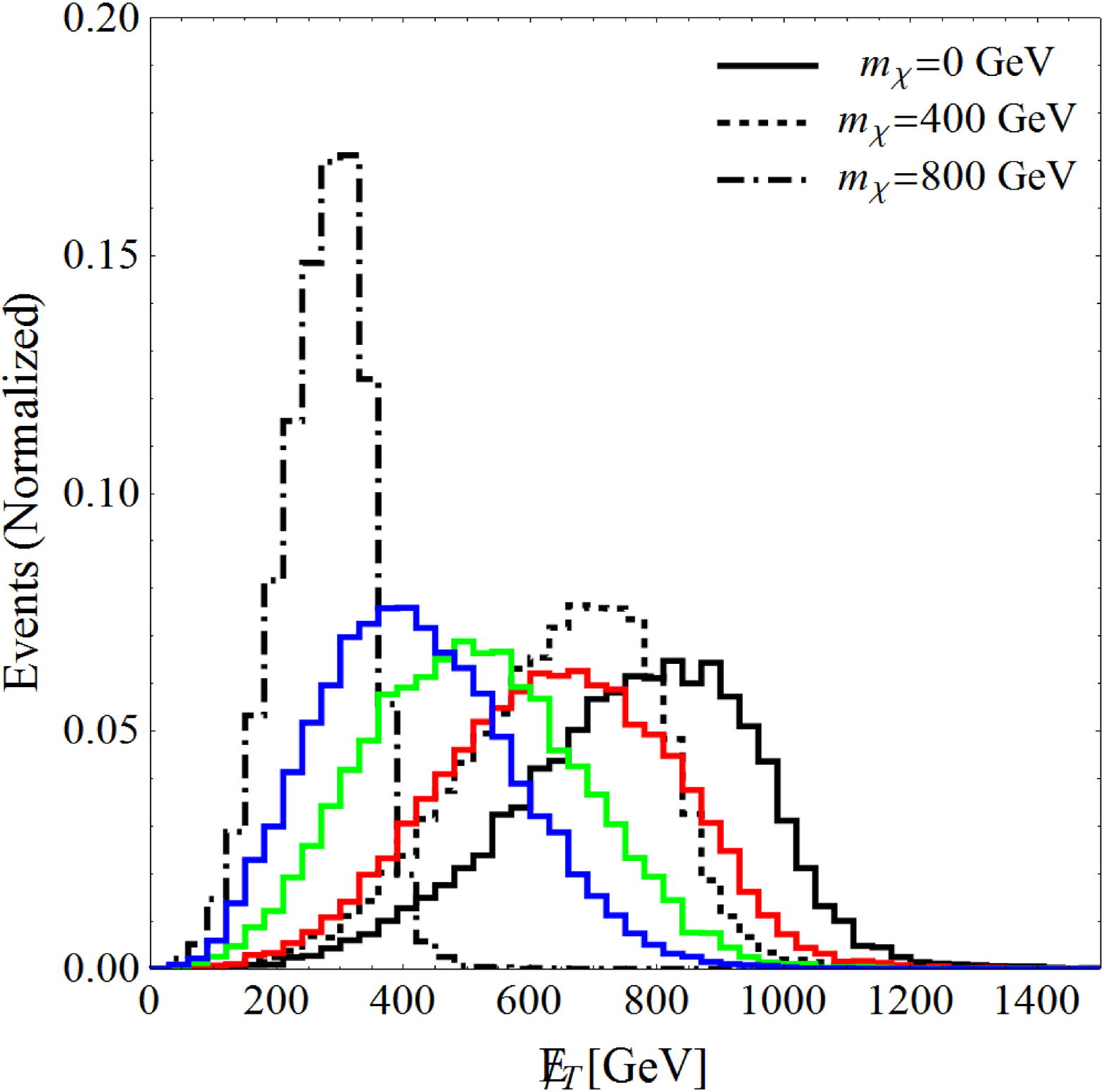}~~
  \epsfxsize 2.15 truein \epsfbox {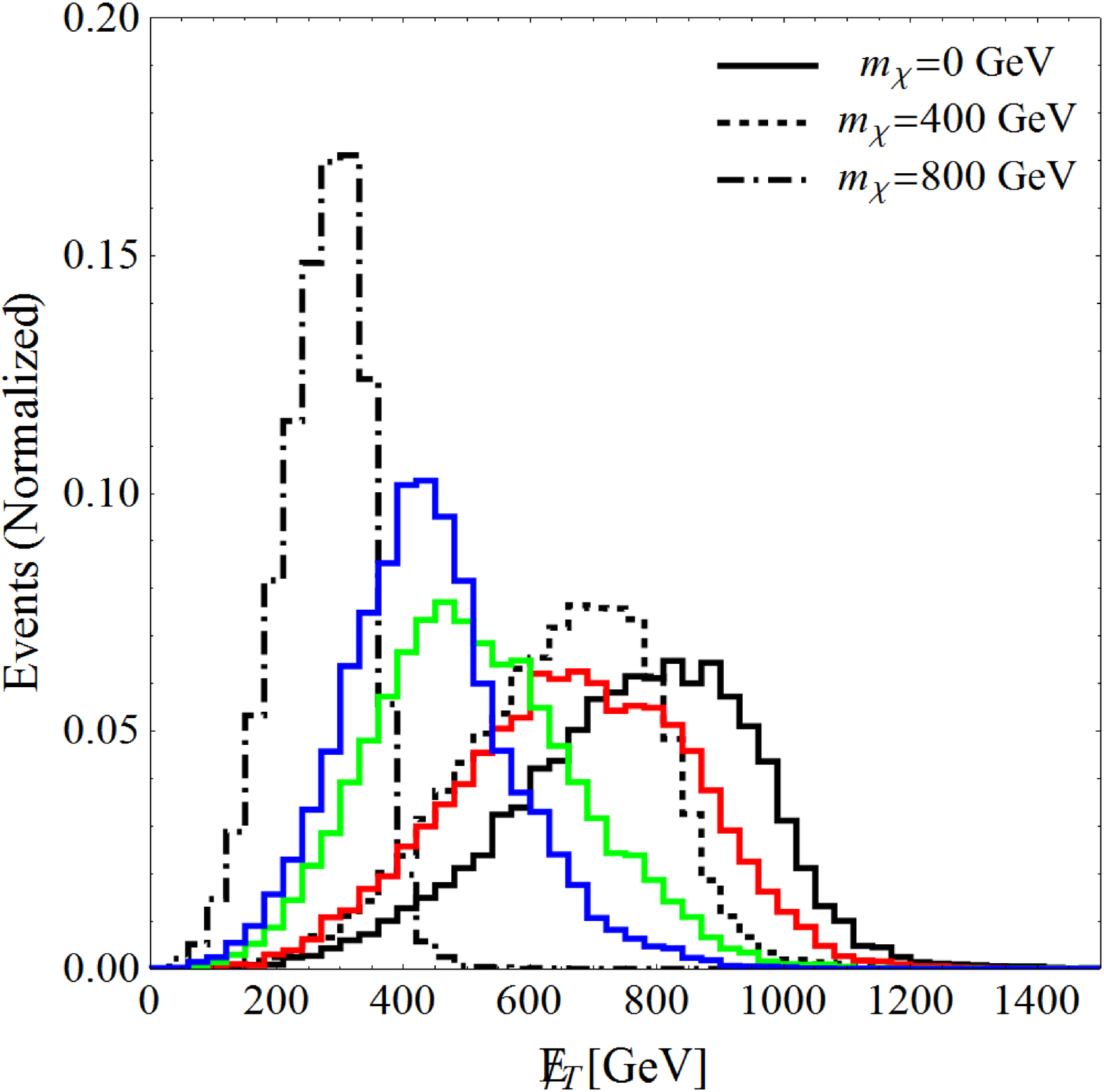}~~
  \epsfxsize 2.15 truein \epsfbox {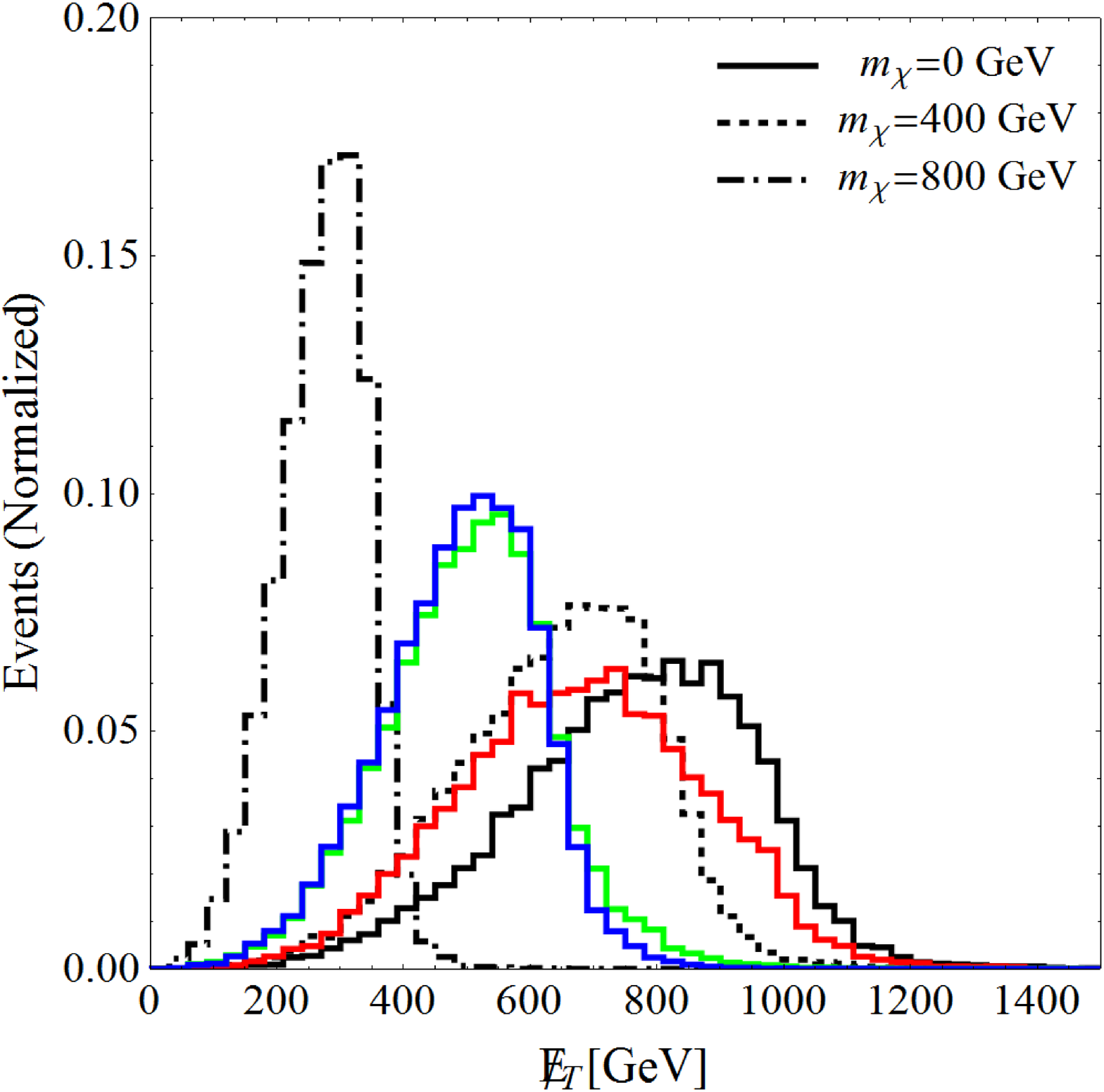}
\end{center}
\caption{ ~Same as Fig.~\protect\ref{fig:METDistsForDDM}, but with the additional 
  cut $\alpha_T>0.55$.}
\label{fig:METDistsForDDMAfterCuts}
\begin{center}
  \epsfxsize 2.15 truein \epsfbox {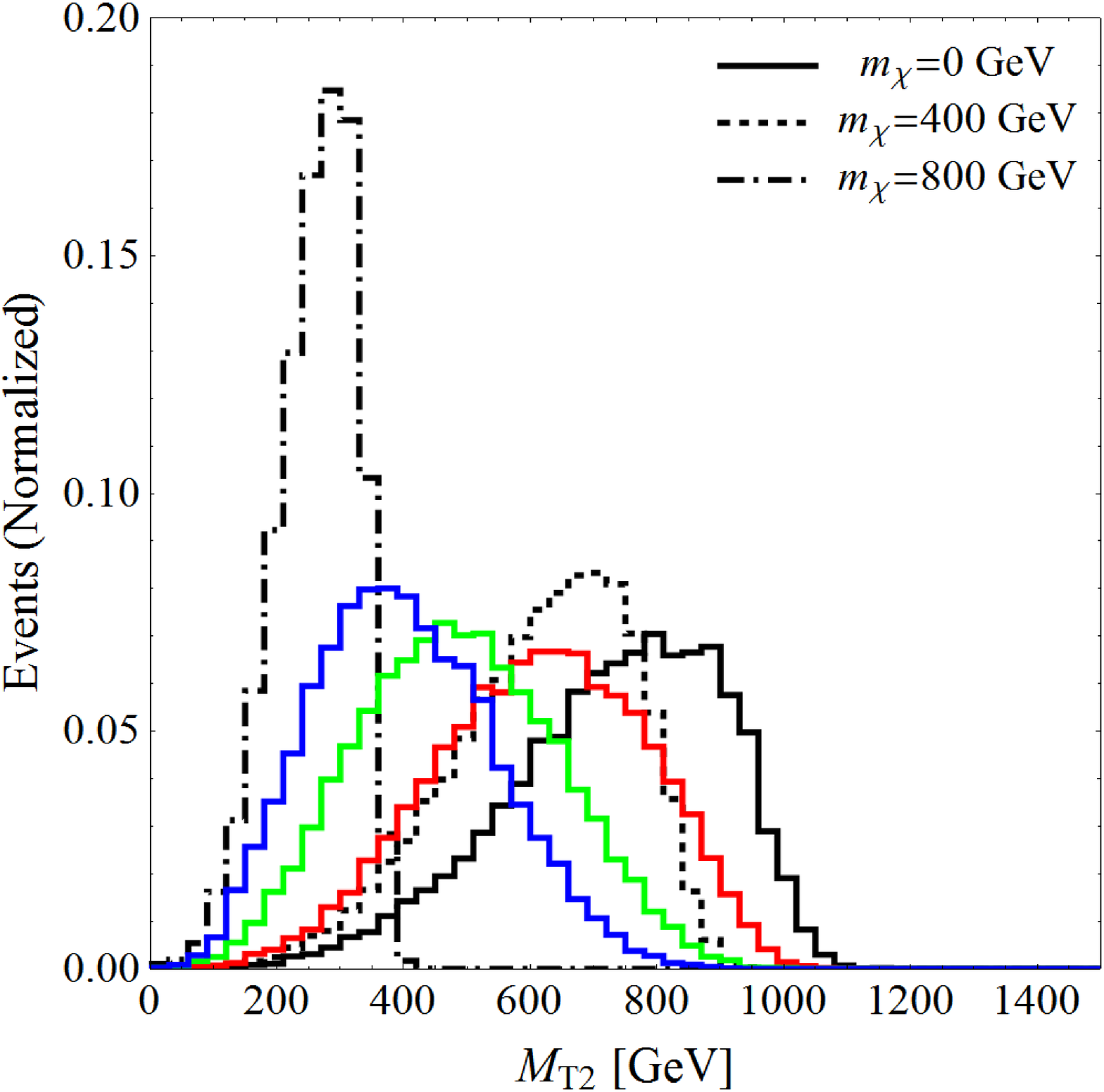}~~
  \epsfxsize 2.15 truein \epsfbox {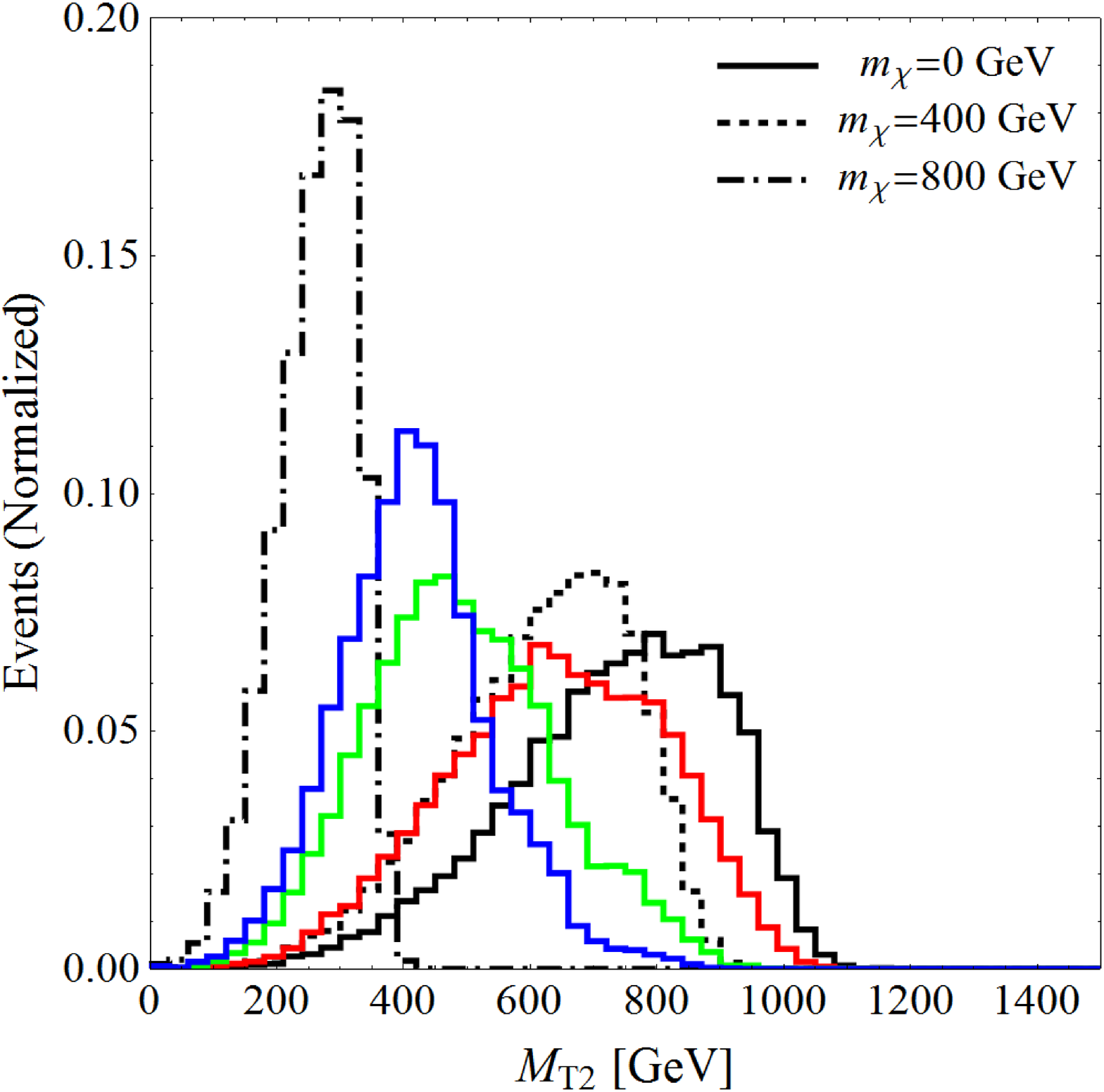}~~
  \epsfxsize 2.15 truein \epsfbox {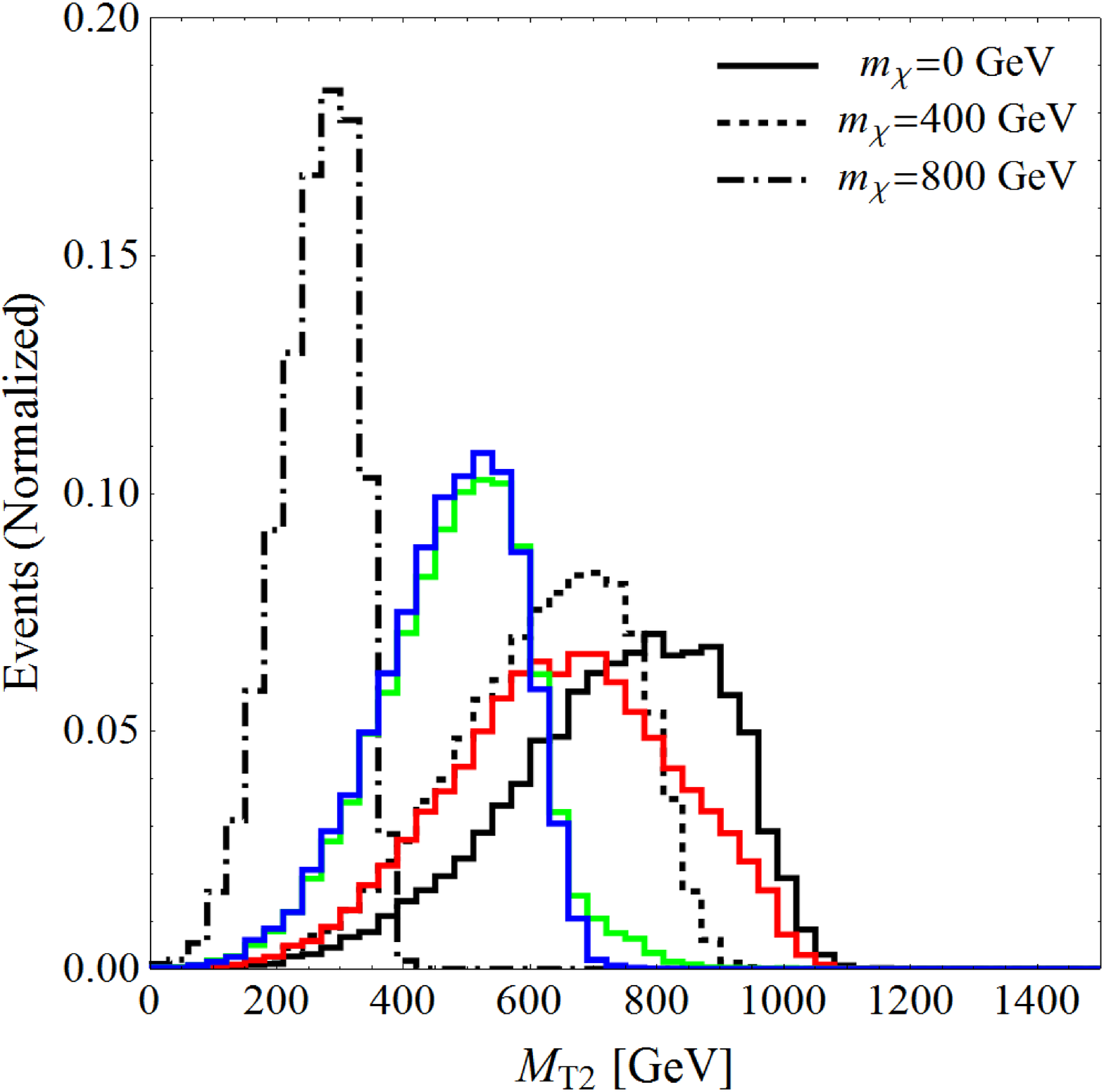}
\end{center}
\caption{ ~Same as in Fig.~\protect\ref{fig:MT2DistsForDDM}, but with the additional 
  cut $\alpha_T>0.55$.}
\label{fig:MT2DistsForDDMAfterCuts}
\end{figure*}

Unfortunately, it is possible for cuts on these variables to significantly affect the
shapes of the distributions we have calculated in Sect.~\ref{sec:VarsDDMvsTDM}.
Such cuts might therefore eliminate not only our backgrounds, but also the ability of
kinematic variables such as $\met$ and $\MT2$ to discriminate between minimal and 
non-minimal dark sectors.  To illustrate this possibility, we can begin with the 
$\met$ and $\MT2$ distributions in Figs.~\ref{fig:METDistsForDDM} and \ref{fig:MT2DistsForDDM} 
and impose a single additional cut $\alpha_T >0.55$.  The results are shown in 
Figs.~\ref{fig:METDistsForDDMAfterCuts} and \ref{fig:MT2DistsForDDMAfterCuts}.
Note that although this cut also results in a substantial reduction in the
signal-event count, our primary focus is on the {\it shapes}\/ of
the distributions.  Thus, the distributions in
Figs.~\ref{fig:METDistsForDDMAfterCuts} and \ref{fig:MT2DistsForDDMAfterCuts} are
likewise normalized so that the area under each distribution is unity.
   
Comparing the distributions in Fig.~\ref{fig:METDistsForDDMAfterCuts}
with those of Fig.~\ref{fig:METDistsForDDM} (or equivalently comparing those of 
Fig.~\ref{fig:MT2DistsForDDMAfterCuts} with those of Fig.~\ref{fig:MT2DistsForDDM}), 
we see that our $\alpha_T$ cut has had a significant impact on the shapes of these
$\met$ and $\MT2$ distributions.  Such cuts can therefore have a significant effect 
on our ability to distinguish minimal from non-minimal dark sectors.
Indeed, this is true even when the variable on which the cut is imposed 
itself displays little sensitivity to the parameters which characterize the dark 
sector --- as we have shown to be the case for $\alpha_T$.

Ultimately, cuts on variables such as $\alpha_T$ and $H_{T_{jj}}$
are able to affect the shapes of $\met$ and $\MT2$ distributions for only
one reason:  {\it there are non-trivial correlations between these two 
groups of variables}\/.  Otherwise, in the absence of such correlations (and 
given sufficient statistics), cuts on these variables would result in a uniform
reduction in signal events across these distributions but leave the overall 
{\it shapes}\/ of these distributions intact.

\begin{figure*}[t!]
\begin{center}
  \epsfxsize 2.15 truein \epsfbox 
    {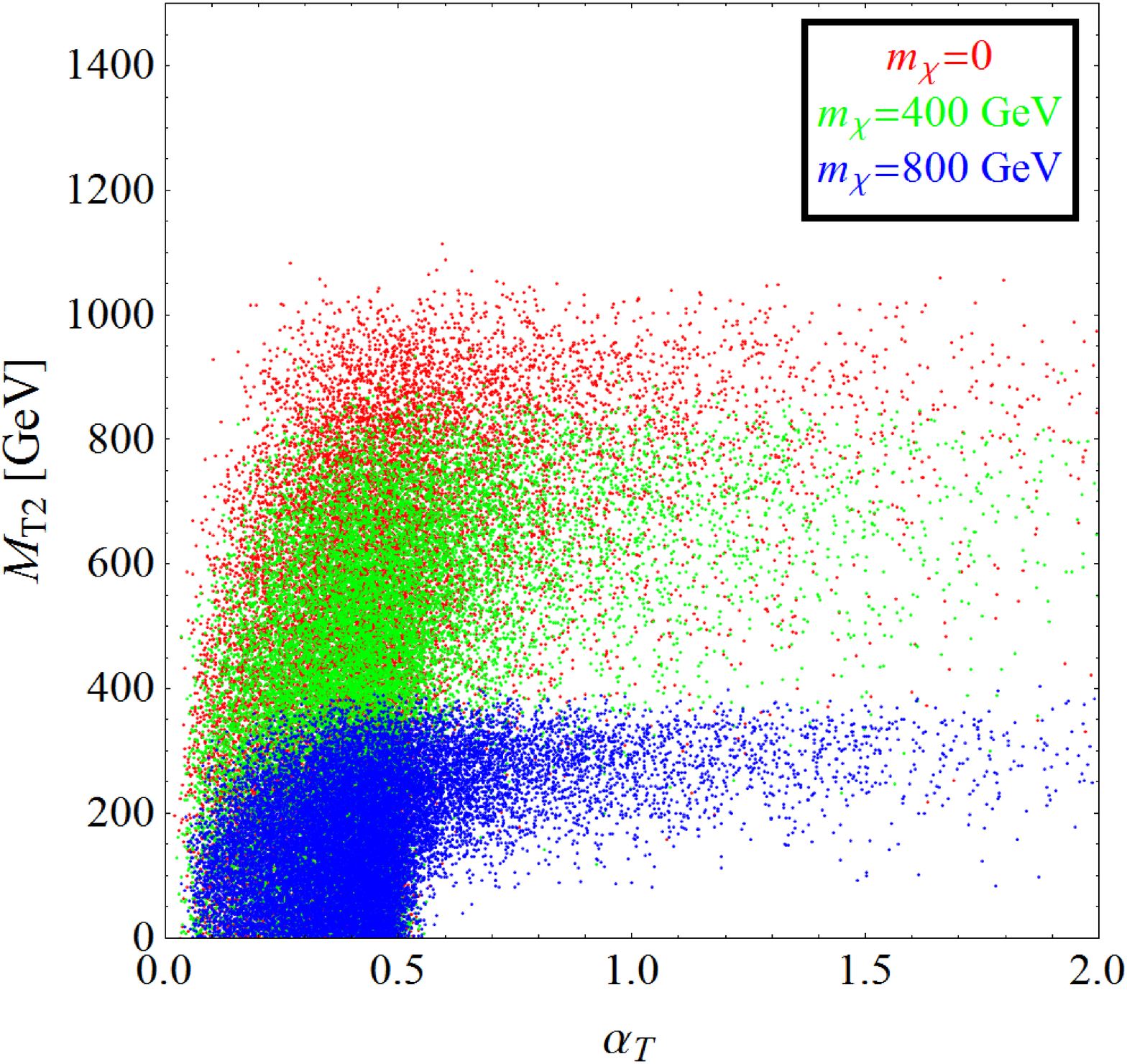}~~
  \epsfxsize 2.15 truein \epsfbox
    {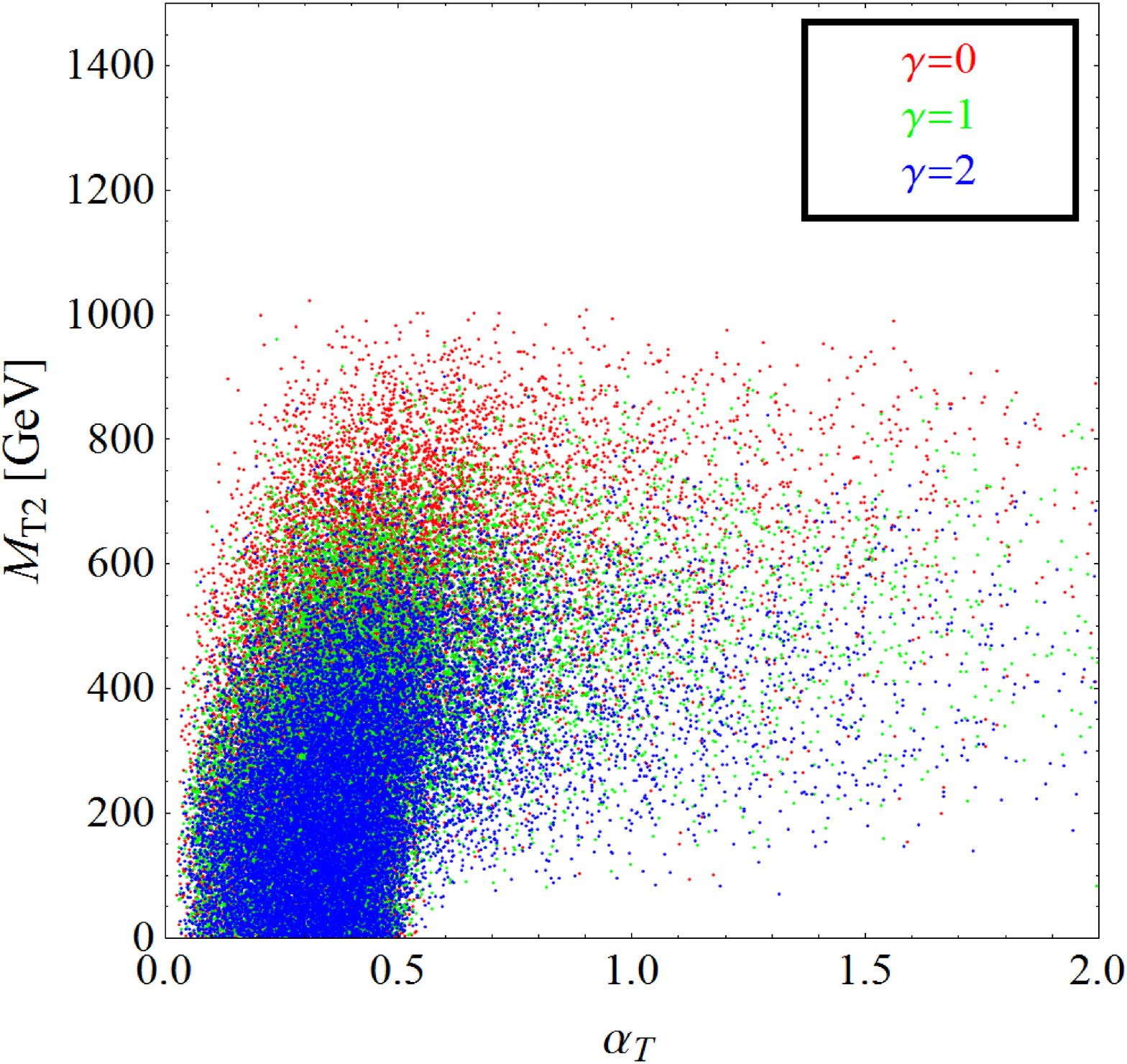}~~
  \epsfxsize 2.15 truein \epsfbox 
    {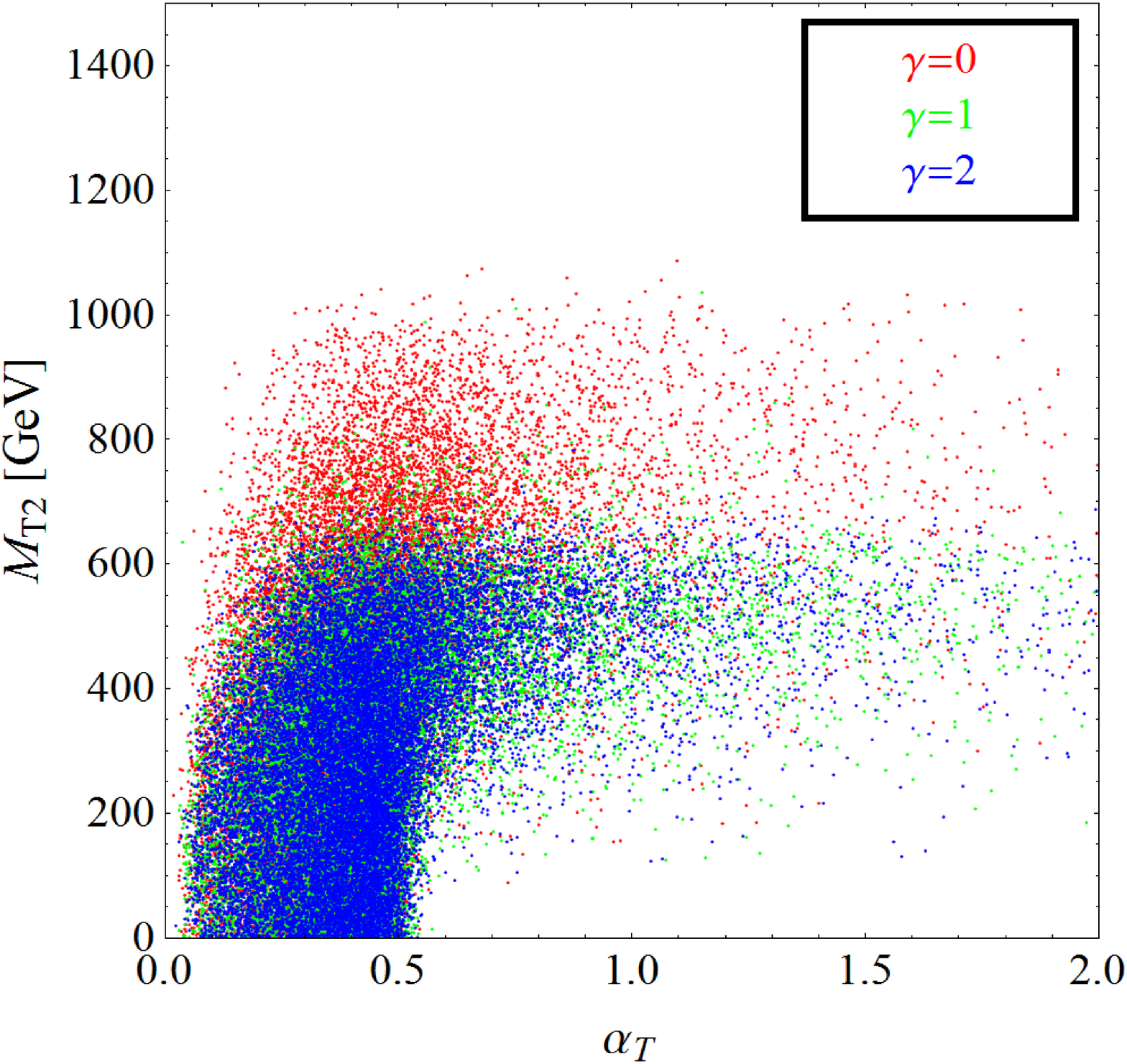}\\
  \epsfxsize 2.15 truein \epsfbox 
    {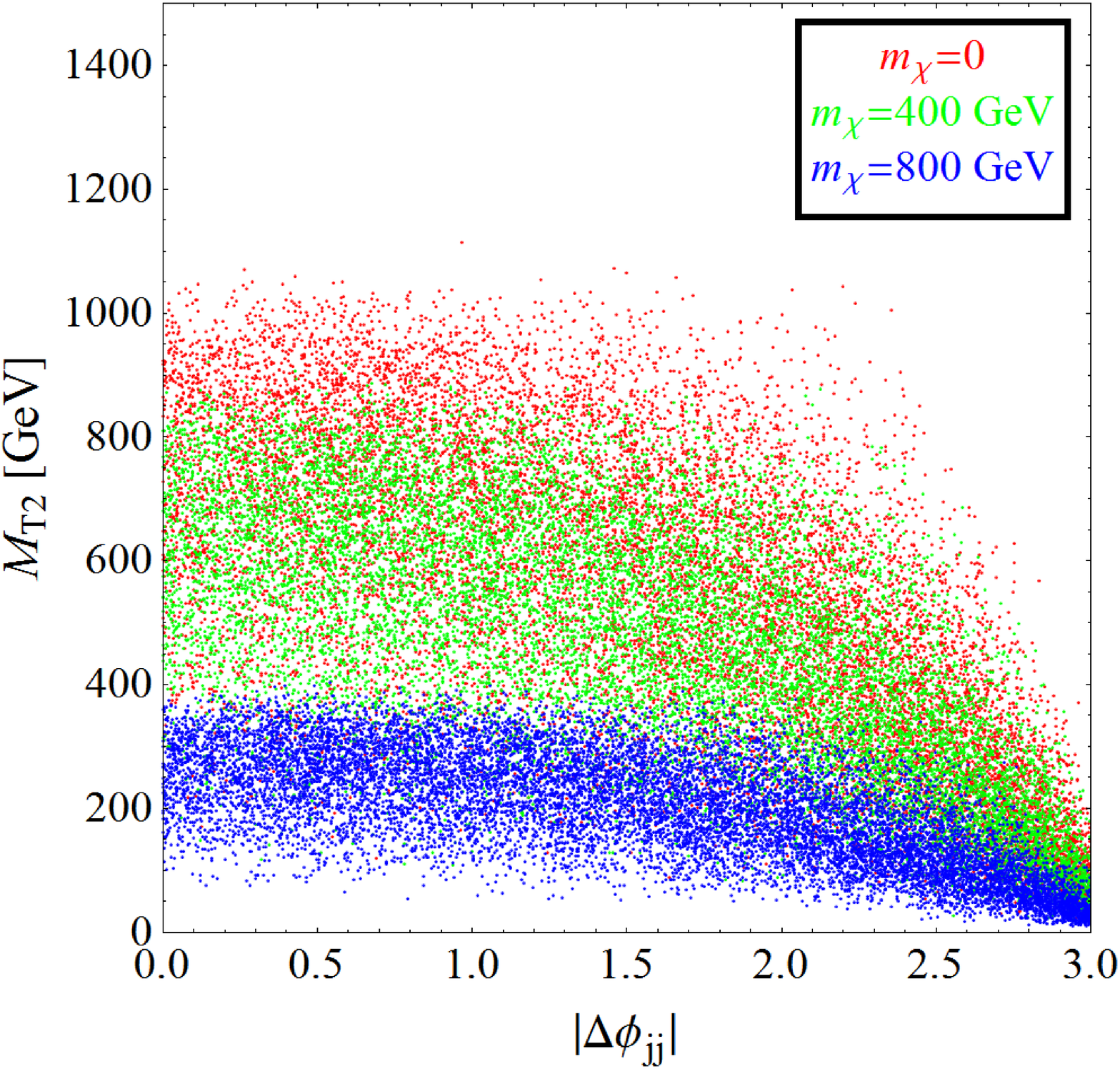}~~
  \epsfxsize 2.15 truein \epsfbox
    {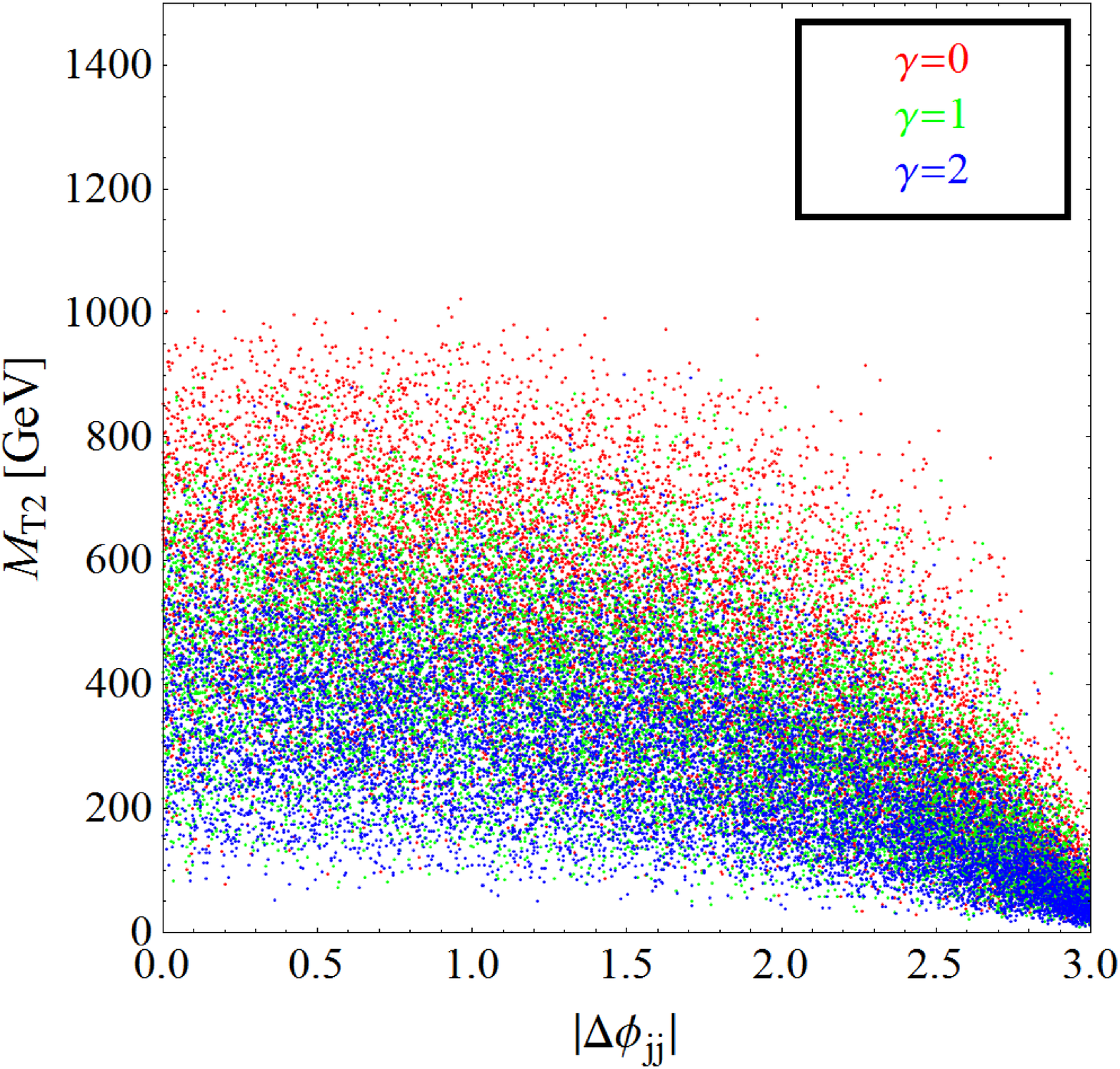}~~
\epsfxsize 2.15 truein \epsfbox 
    {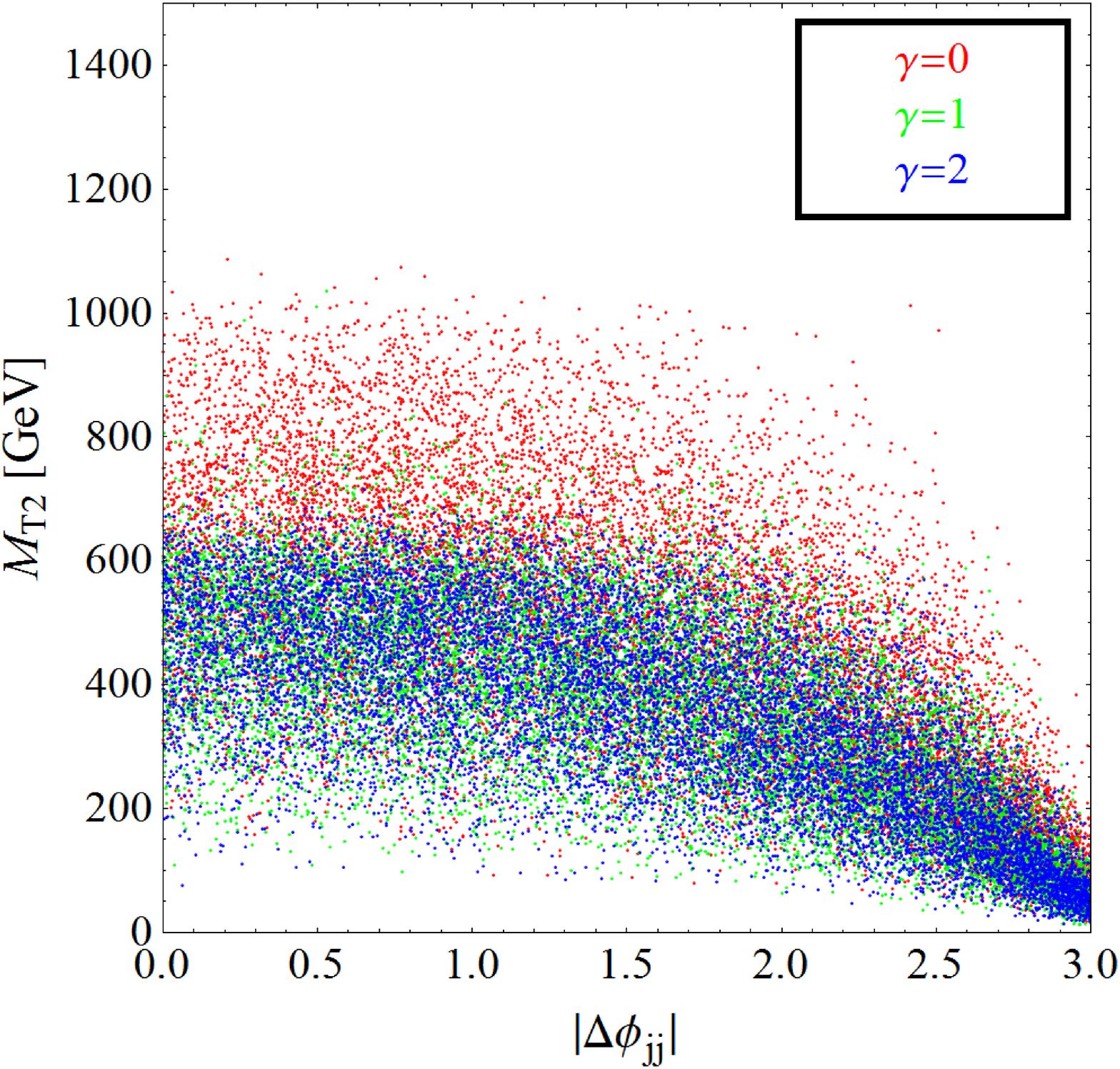}
  \epsfxsize 2.15 truein \epsfbox 
    {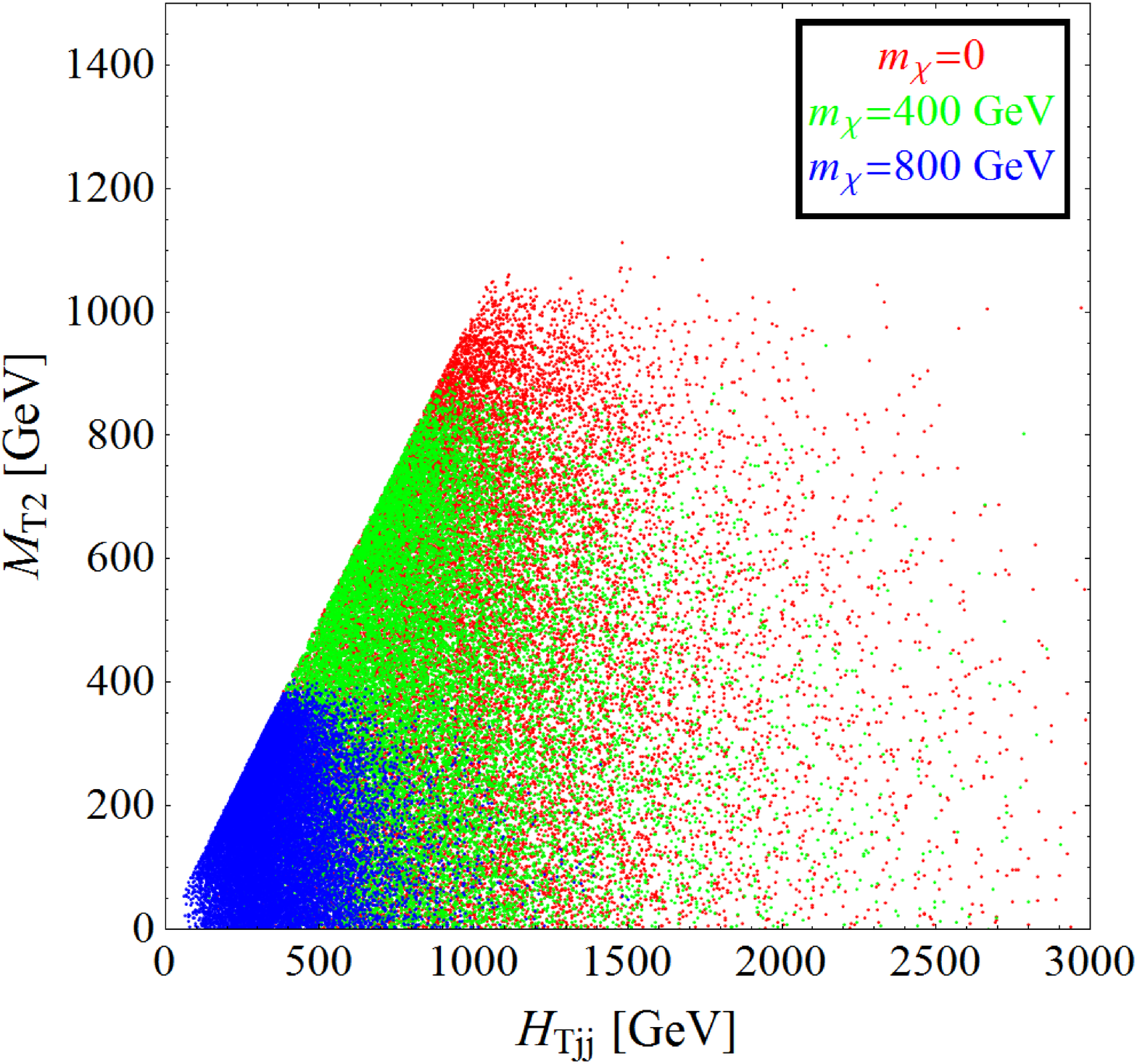}~~
  \epsfxsize 2.15 truein \epsfbox
    {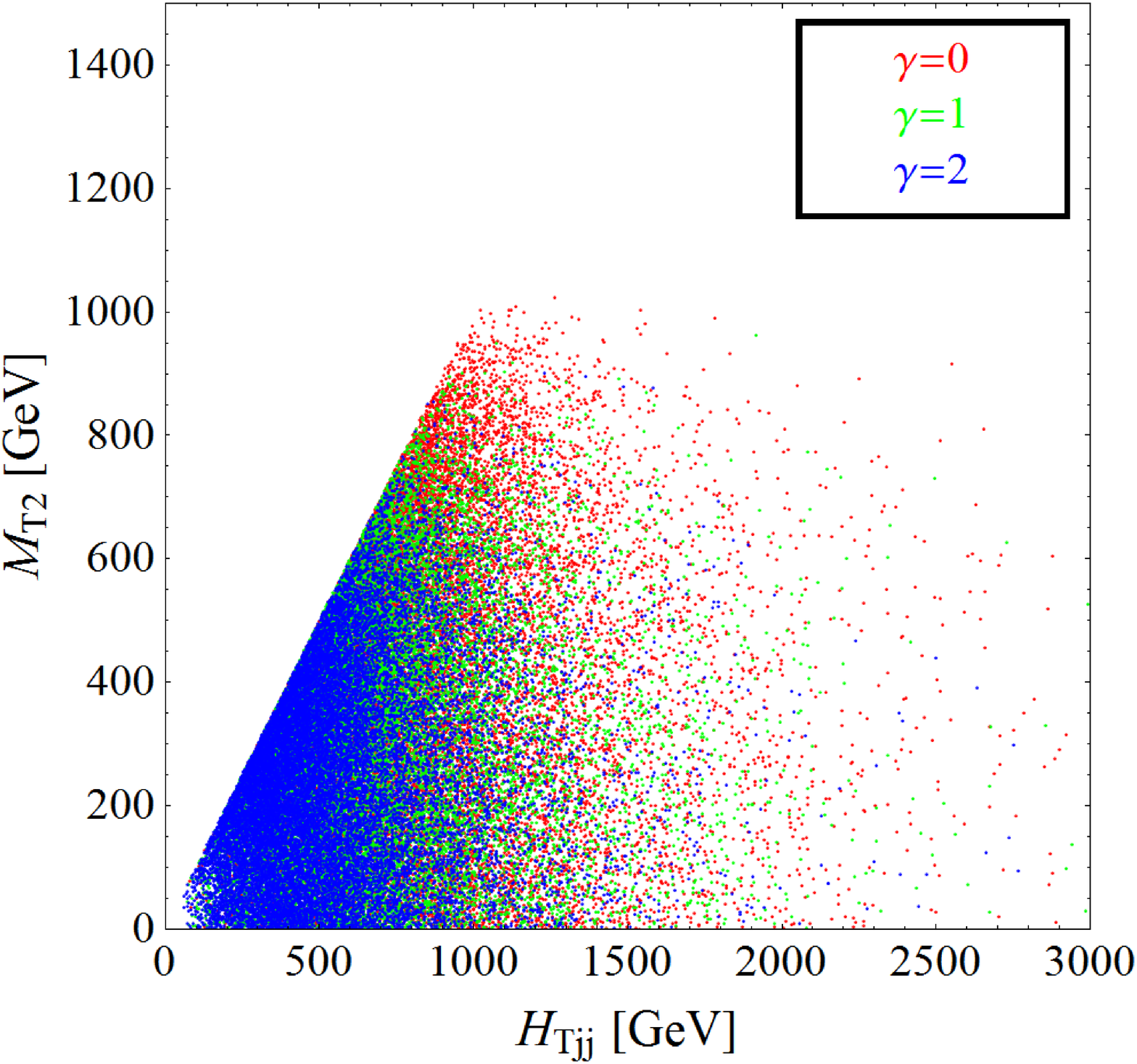}~~
\epsfxsize 2.15 truein \epsfbox 
    {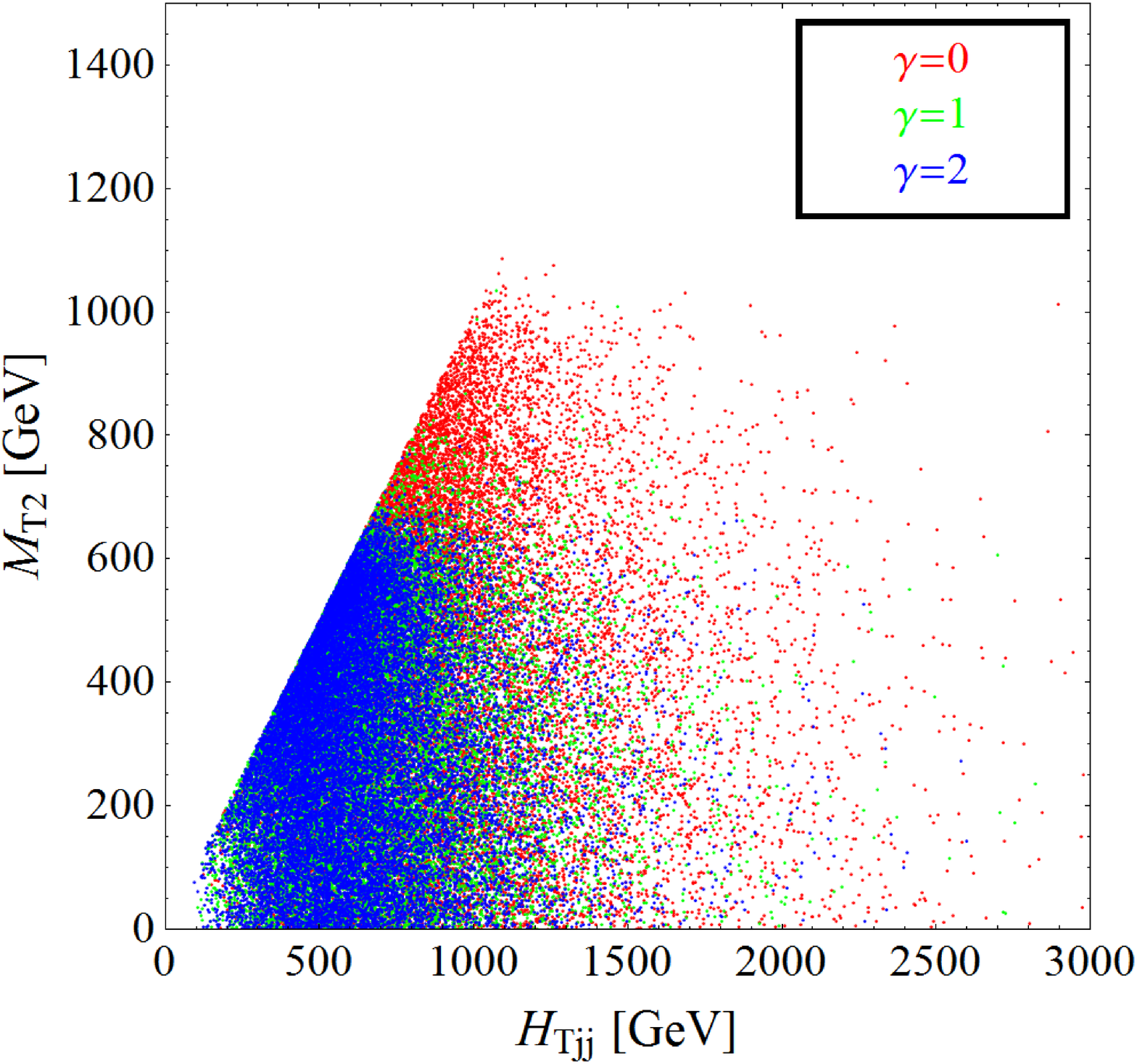}
\end{center}
\caption{ ~Scatter plots illustrating the correlations between $\MT2$ and 
  the selection variables $\alpha_T$ (top row), $|\Delta\phi_{jj}|$ 
  (center row), and $H_{T_{jj}}$ (bottom row) for a trial mass 
  $\widetilde{m} = 0$.  The left panel in
  each row shows the results for traditional dark-matter models with
  $m_\chi = 0$ (red), $m_\chi = 400$~GeV (green), and $m_\chi = 800$~GeV (blue).  
  The center panel in each row shows the results for three DDM models, with 
  $m_0 = 100$~GeV, $\Delta m = 50$~GeV, $m_\phi = 1$~TeV, and
  $\gamma = 0$ (red), $\gamma = 1$ (green), and $\gamma = 2$ (blue).  The
  right panel in each row shows the corresponding results for a DDM model with 
  $\Delta m = 500$~GeV and all other parameters unchanged.
\label{fig:ScatterPlotsMT2}}
\end{figure*}

In order to explore this issue further, we turn to directly examine the
correlations between the variables $\lbrace \alpha_T, |\Delta\phi_{jj}|, H_{T_{jj}}\rbrace$  --- 
which are important for removing backgrounds and extracting signals ---
and the variables $\lbrace \met, \MT2\rbrace$ --- which are also important for distinguishing
between minimal and non-minimal dark sectors.  The correlations between this former set of
variables and $\MT2$ (with a trial mass $\widetilde{m} = 0$) are illustrated in the 
scatter plots displayed in Fig.~\ref{fig:ScatterPlotsMT2} for a benchmark set of traditional 
dark-matter models (left column) and DDM ensembles (center and right columns).  In each of these 
scatter plots, we display several sets of data points associated with these different benchmark 
models.  Each data point corresponds to a single event chosen randomly from the Monte-Carlo data 
sample for that model: its color indicates the model with which it is associated and
its coordinates indicate the values $x$ and $y$ of the two kinematic variables $X$ and $Y$ of
interest for the corresponding event.  Thus, the density of points within the region 
$(x,y)$ to $(x + \delta x, y + \delta y)$ indicates the relative likelihood of values
for $X$ and $Y$ within that range occurring in combination.  A uniform density of points 
throughout a particular panel would imply that the variables are essentially uncorrelated.

Clearly, the results shown in Fig.~\ref{fig:ScatterPlotsMT2} indicate not only that 
$\MT2$ and the variables $\lbrace \alpha_T, |\Delta\phi_{jj}|, H_{T_{jj}}\rbrace$ are 
correlated in interesting, non-trivial ways, but also that these 
correlations often depend sensitively on the masses and couplings of the dark-sector
particles!  For example, we observe from the top left panel of this figure  
that there is far less overlap among the data-point distributions for traditional 
dark-matter models with different $m_\chi$ within the region of parameter space in 
which $\alpha_T > 0.55$ than there is within the region in which $\alpha_T < 0.55$.
Therefore, while a cut on $\alpha_T$ of this magnitude does significantly reduce 
the total number of signal events, this effect is offset at least in part by the
fact that the $\MT2$ distributions of the surviving data points for different $m_\chi$  
are significantly more segregated from one another after the cut than before.  Indeed,
the way in which $\alpha_T$ and $\MT2$ are correlated makes $\alpha_T$ a particularly 
effective selection variable: not only are cuts on $\alpha_T$ effective in reducing 
SM backgrounds, but they also serve to {\it amplify}\/ distinctions between the shapes 
of the $\MT2$ distributions associated with different dark-matter models!  Indeed, as
we shall see in Sect.~\ref{sec:results}, this effect more than compensates for the loss 
in signal-event count that arises for a threshold cut $\alpha_T^{\mathrm{min}}$ on $\alpha_T$ 
of the order necessary to effectively eliminate the substantial QCD background.  

It is likewise evident from Fig.~\ref{fig:ScatterPlotsMT2} that similar correlations
exist between $|\Delta\phi_{jj}|$ and $\MT2$.  However, cuts on $\alpha_T$ and 
$|\Delta\phi_{jj}|$ are to a large extent redundant, since both variables essentially
reflect the degree to which the leading two jets in a given event are back-to-back.  
Moreover, we note that cuts on $\alpha_T$ are typically found to be significantly more 
effective in reducing SM backgrounds than cuts on $|\Delta\phi_{jj}|$.  For these reasons,
we henceforth focus on $\alpha_T$, but we note that results similar to those we obtain in
this study could in principle be obtained by imposing cuts on $|\Delta\phi_{jj}|$ 
rather than $\alpha_T$.

By contrast, Fig.~\ref{fig:ScatterPlotsMT2} reveals that $H_{T_{jj}}$ and $\MT2$ are 
correlated in such a way that imposing a substantial minimum cut on $H_{T_{jj}}$ 
distorts $\MT2$ distributions in a reliable but far less advantageous manner.    
In particular, we observe from the bottom left panel of this figure 
that a minimum cut of $H_{T_{jj}}$ results in a far more severe reduction in
signal events for traditional dark-matter models with large $m_\chi$ than for those
with small $m_\chi$.  This in turn implies that in DDM scenarios and other theories 
involving multiple invisible particles, information about the heavier particles
tends to be washed out by the application of such a cut.  It therefore follows that
DDM ensembles with extremely large values of the coupling index $\gamma$ will be
somewhat more difficult to distinguish on the basis of their $\MT2$ distributions, 
since the branching fractions of the parent particle $\phi$ to the heavier $\chi_i$ 
are comparatively large in this case.

We note that while the results shown in Fig.~\ref{fig:ScatterPlotsMT2} correspond
to the choice of $\widetilde{m} =0$, we find that the corresponding results for other choices of 
this trial mass exhibit the same qualitative features.  However, as discussed in 
Sect.~\ref{sec:VarsDDMvsTDM}, the window of possible $\MT2$ values narrows as 
$\widetilde{m}$ increases.  In general, this narrowing 
results in a greater degree of overlap among the the data-point distributions associated 
with different dark-matter models and consequently makes distinguishing among such models 
more difficult.  We also note that in both the $\Delta m \rightarrow \infty$ and 
$\gamma \rightarrow -\infty$ limits, the data-point distribution associated with a DDM 
ensemble reduces to that associated with a single dark-matter particle with $m_\chi = m_0$, 
as expected.     
 
Since have seen in Sect.~\ref{sec:VarsDDMvsTDM} that the shape of the $\met$ distribution 
is also sensitive to the structure of the dark sector, it is interesting to examine the 
correlations between $\met$ and 
the variables $\lbrace \alpha_T, |\Delta\phi_{jj}|, H_{T_{jj}}\rbrace$ as well.
We find that each of these variables turns out to be correlated with $\met$ in a manner 
extremely similar to that in which it is correlated with $\MT2$.  Indeed, 
the corresponding scatter plots are qualitatively so similar to those shown in 
Fig.~\ref{fig:ScatterPlotsMT2} that we refrain from reproducing them here.
As we shall see in Sect.~\ref{sec:results}, these correlations likewise imply that 
imposing cuts on variables such as $\alpha_T$, $|\Delta\phi_{jj}|$, and $H_{T_{jj}}$ 
can have a significant effect on the distributions of both $\met$ and $\MT2$. 


\section{Comparing Correlations:  Balancing Signal Extraction Against 
Dark-Sector Resolution\label{sec:results}}


As we have seen, the existence of non-trivial correlations between the
variables $\lbrace \alpha_T, |\Delta\phi_{jj}|, H_{T_{jj}}\rbrace$ and the variables 
$\lbrace \met,\MT2 \rbrace$ implies that cuts on variables in the first set 
will distort the shapes of the kinematic distributions associated with variables 
in the second set.  This is generically true for both traditional single-component 
dark sectors and non-minimal dark sectors.  However, these correlations
are ultimately a reflection of fundamental kinematic relationships between the 
masses, energies, and momenta of the particles involved in the production and 
decay processes associated with any particular event.  This means that the way in 
which kinetic variables are correlated --- and therefore the effect that a cut on
one such variable will have on the distribution of another --- is itself 
dependent on the properties of the dark-sector particles.  

As a result, the degree to which we can ultimately exploit the distributions associated 
with the $\lbrace \met,\MT2 \rbrace$ variables in order to distinguish between 
any two dark-matter models --- and, by extension, between minimal and non-minimal 
dark sectors --- rests upon our understanding of the correlations that exist for   
those models.  Indeed, in order to assess the degree to which dark-sector 
non-minimality can be distinguished at the LHC, we must compare the effects of the 
correlations which arise in non-minimal dark sectors with the effects of the 
correlations which arise in minimal dark sectors.  More specifically, for the case at 
hand, we must ultimately {\it compare}\/ the correlations illustrated in the second 
and third columns of Fig.~\ref{fig:ScatterPlotsMT2} with those illustrated in the first 
column of Fig.~\ref{fig:ScatterPlotsMT2}.

\begin{figure*}[t!]
\begin{center}
  \epsfxsize 2.370 truein 
    \epsfbox {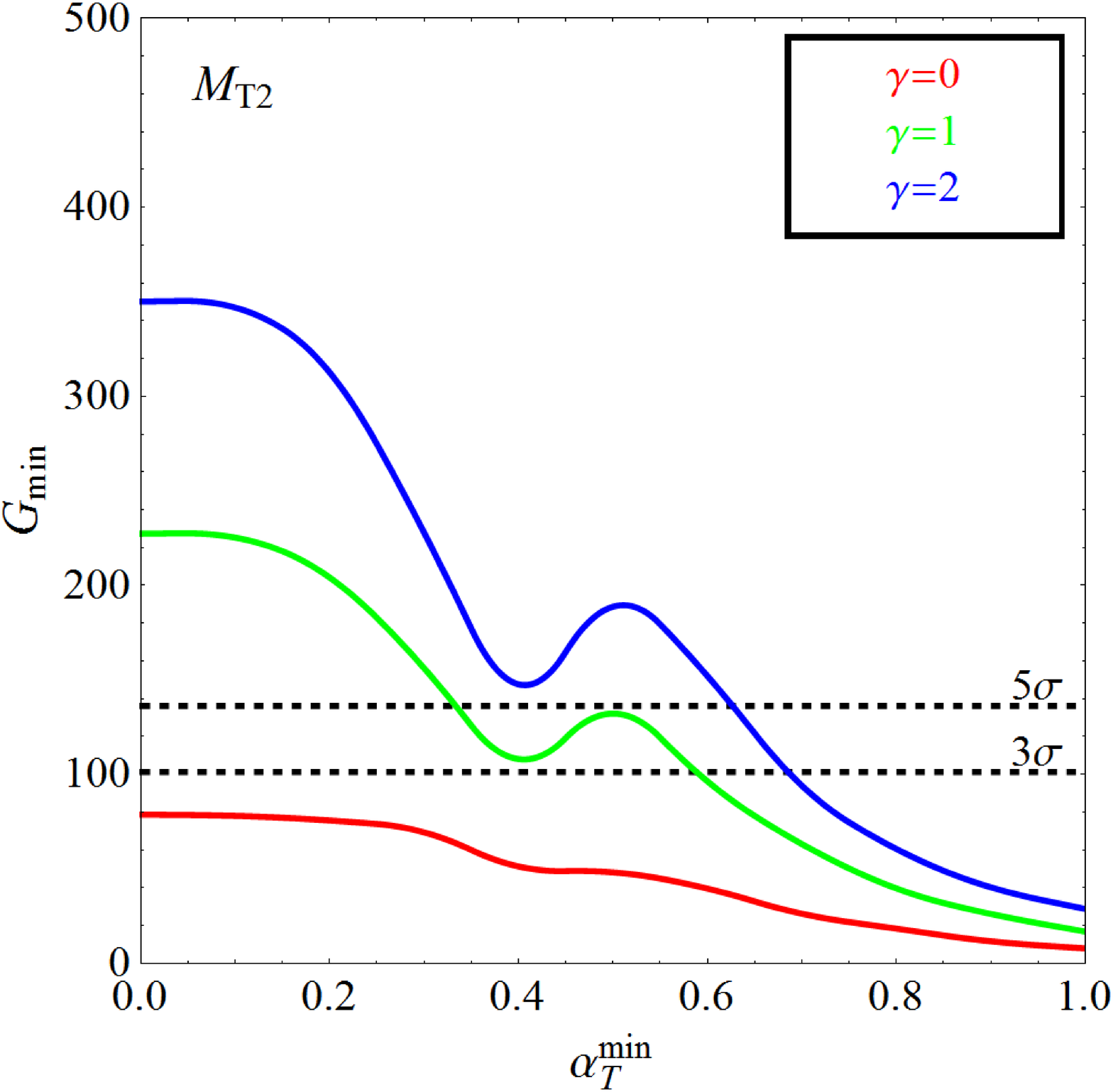}~~~~
  \epsfxsize 2.3 truein 
    \epsfbox {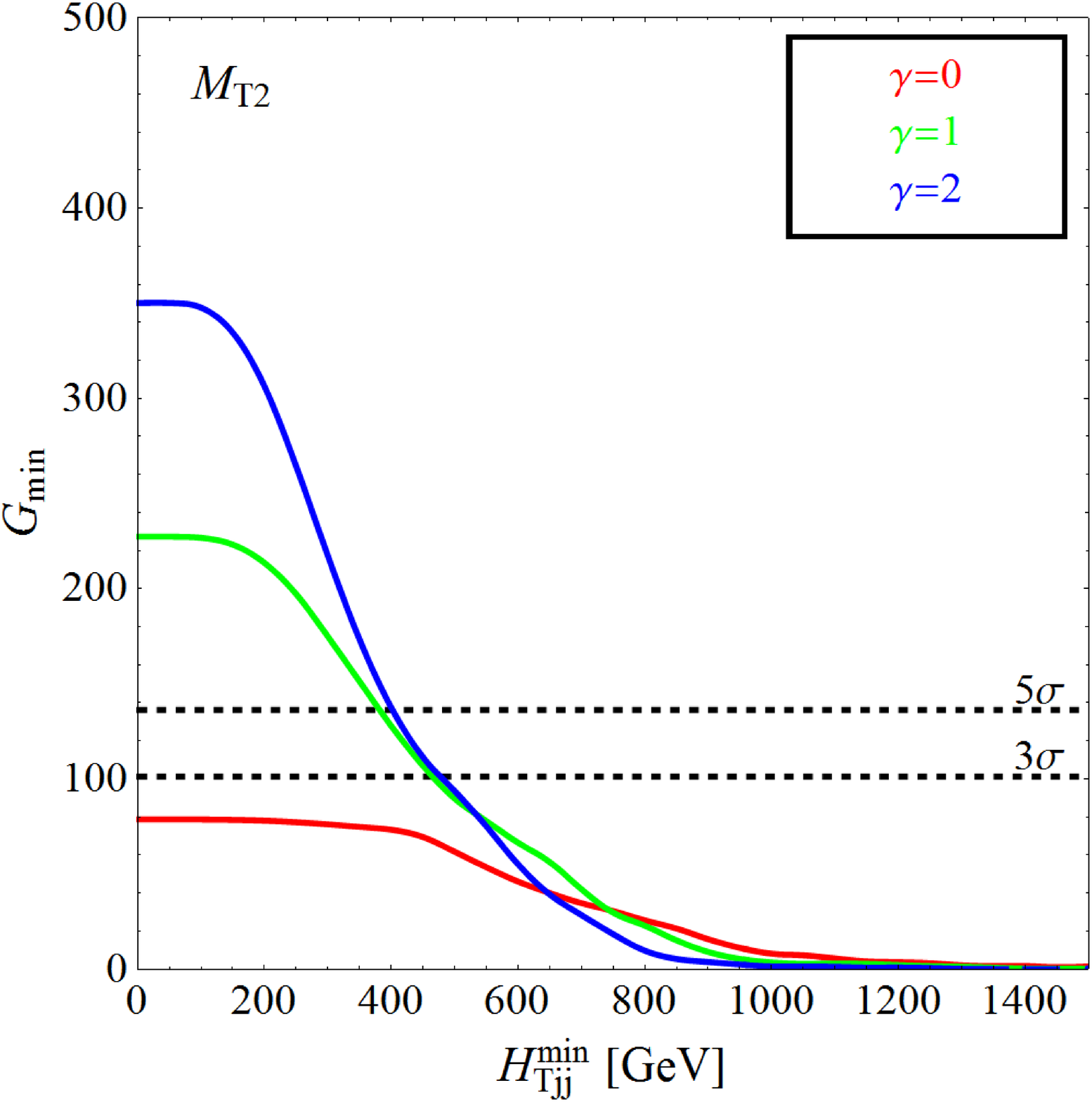}\\
  \epsfxsize 2.370 truein 
    \epsfbox {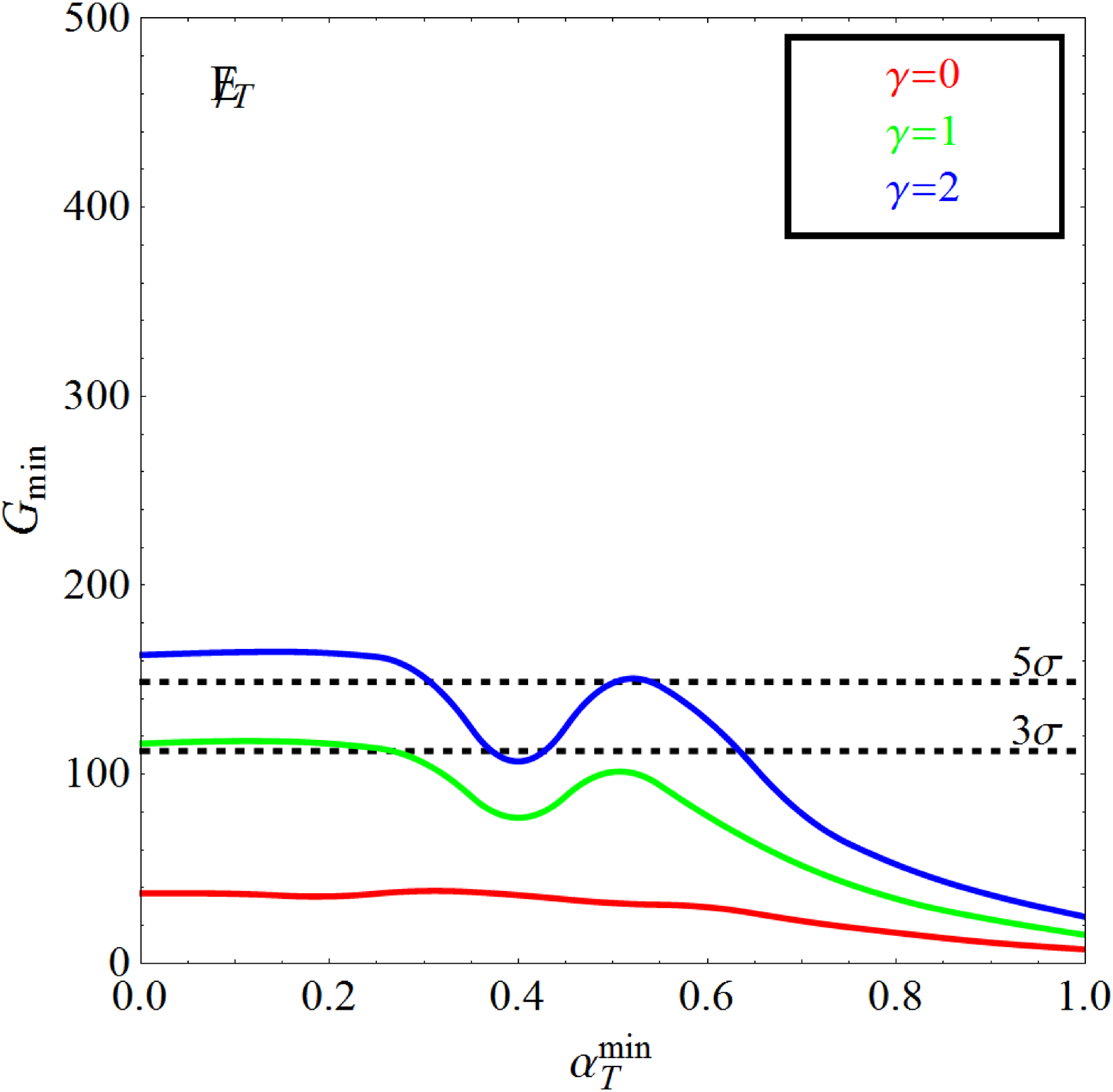}~~~~
  \epsfxsize 2.3 truein 
    \epsfbox {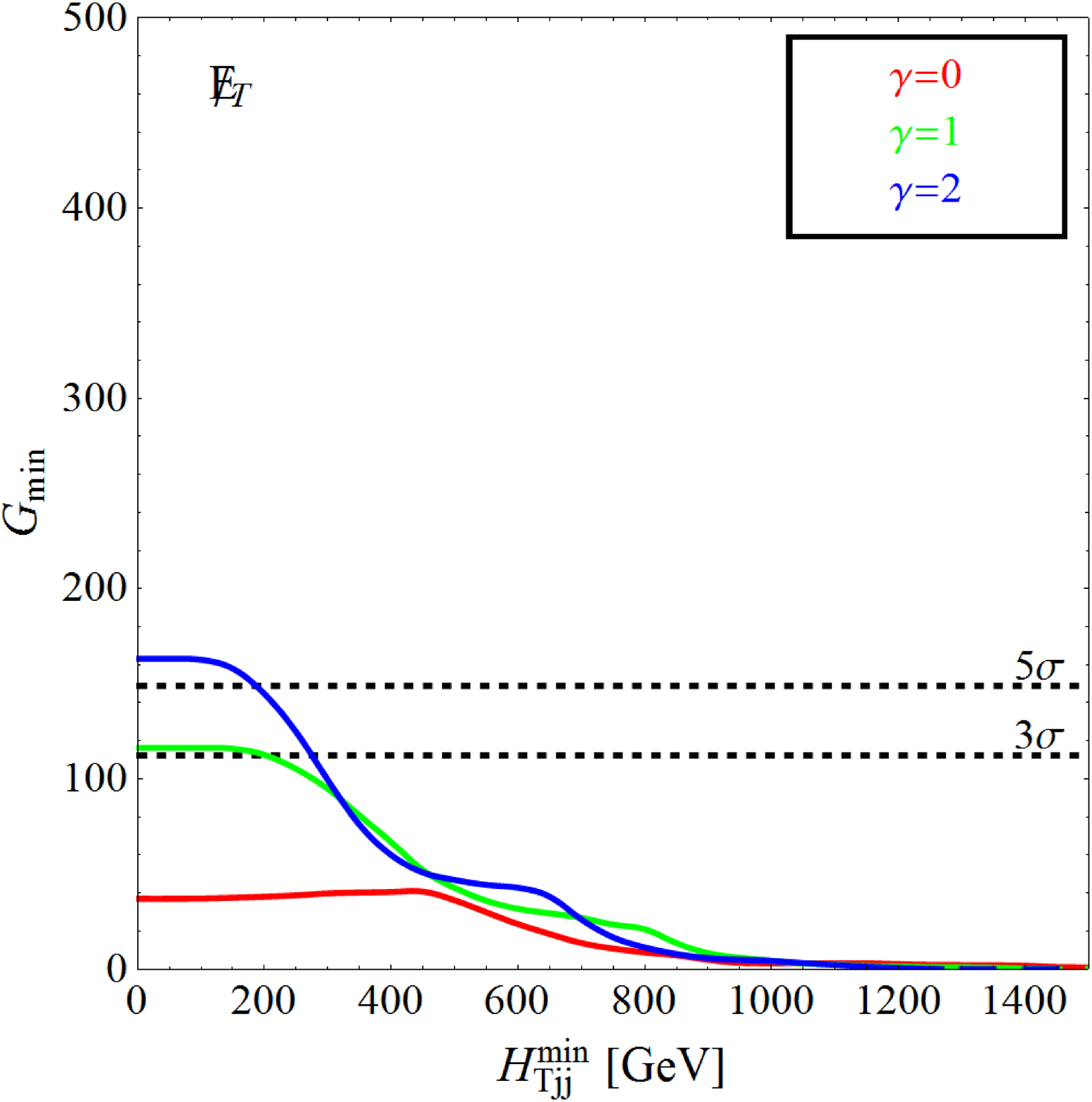}
\end{center}
\caption{ ~The value of the statistic $G_{\mathrm{min}}$ for the $\MT2$ distributions 
(top row) and $\met$ distributions (bottom row) associated with several DDM ensembles 
as a function of the minimum cut $\alpha_T^{\mathrm{min}}$ imposed on $\alpha_T$ 
(left column) or the minimum cut $H_{T_{jj}}^{\mathrm{min}}$ imposed on $H_{T_{jj}}$ 
(right column).  In each case, the precuts in Sect.~\protect\ref{sec:CalculatingDistributions} 
are the only other cuts imposed on the data.  The curves shown in all panels correspond to the 
parameter choices $m_0 = 100$~GeV, $\Delta m = 50$~GeV, and $\delta =1$, and the $\MT2$ curves 
correspond to a trial mass $\widetilde{m} = 0$.  Further details are discussed in the text.  
\label{fig:GminCurves}}
\end{figure*}

In order to make this comparison, we require a method of quantifying the degree to which
the expected distribution for any particular collider variable associated with a given DDM 
ensemble is distinct from the distributions associated with traditional dark-matter 
candidates in general.  To do this, we shall adopt a procedure similar to that employed in 
Ref.~\cite{DDMLHC1}.  In particular, we survey over a variety of traditional dark-matter
candidates, each characterized by a different value of $m_\chi$, and compare the distribution 
of that variable with the distribution obtained for the DDM ensemble of interest.  
We include in our survey values of $m_\chi$ ranging from $m_\chi = 0$ to $m_\chi = m_\phi$ 
at intervals of $100$~GeV.  For each value of $m_\chi$, we assess the degree of distinctiveness 
between the two distributions by 
computing the goodness-of-fit statistic $G(m_\chi) \equiv -2\ln \lambda(m_\chi)$, where 
$\lambda(m_\chi)$ is the ratio of the likelihood functions for the two distributions.
For binned data in which the number of events in each bin is independent of 
the number of events in every other bin, $G(m_\chi)$ takes the form   
\begin{equation}
  G(m_\chi) ~=~ 2 \sum_{k=1}^{N} \left[ \mu_k(m_\chi) - n_k +
    n_k \ln \left(\frac{n_k}{\mu_k(m_\chi)}\right) \right]~,
  \label{eq:DefOfGStatistic}
\end{equation} 
where the index $k$ labels the bin, where $n_k$ is the expected 
population of events in bin $k$ in the DDM model, and where $\mu_k(m_\chi)$ is the 
expected population of events in bin $k$ in the traditional dark-matter model to
which this DDM model is being compared.  We take the minimum value 
\begin{equation}
  G_{\mathrm{min}} ~\equiv~ \underset{m_\chi}{\min} \big\{ G(m_\chi)\big\}
  \label{eq:DefOfGmax}
\end{equation}
from among the $G(m_\chi)$ obtained in our survey over $m_\chi$ as our final measure of 
the distinctiveness of the distribution associated with the DDM ensemble.  Moreover, 
for the case in which the expected population of events in each bin is sufficiently large,   
the $G_{\mathrm{min}}$ statistic follows a $\chi^2$ distribution with 
$N-1$ degrees of freedom.  We can therefore estimate the statistical significance 
of our results by comparing the value of $G_{\mathrm{min}}$ to such a $\chi^2$ distribution 
in order to obtain a $p$-value.  We then take the number of standard deviations away from 
the mean to which this $p$-value would correspond for a Gaussian distribution as an estimate 
of the statistical significance of differentiation --- \ie, the statistical 
significance with which the signal distribution can be claimed to differ 
from the expected distributions for traditional dark-matter candidates.   
We note that in principle one could also incorporate a broader class of traditional
dark-matter models with other particle properties and coupling structures into this
survey; however, the inclusion of such additional models in our analysis will not 
significantly impact our results. 
  
One subtlety which arises in quantifying the discrepancy between different
kinematic distributions is that the reliability of most goodness-of-fit statistics
breaks down in cases in which there are bins for which the expected number of events 
in the reference model is small or zero.  For example, the $G(m_\chi)$ statistic defined 
in Eq.~(\ref{eq:DefOfGStatistic}) is infinite if the expected number of events in one or
more bins is zero in the traditional dark-matter model to which the DDM model is being 
compared.  In order to address this issue, the event count in each bin
in a given background-event distribution for which $\mu_k(m_\chi) < 3$ is treated as
if it were $\mu_k(m_\chi) = 3$.  This is the event count which corresponds to the 
95\%~C.L. upper limit on the expected number of events for data which follow 
a Possion distribution in the case in which no events are observed~\cite{PDG}.  
Likewise, the event count in each bin in a given signal-event distribution for which 
$n_k < 3$ is treated as if it were $n_k = 3$.  Note that while the normalization 
of each distribution is in principle fixed by the measurement of the total number
of signal events, this procedure for treating low-statistics bins does not necessarily 
preserve the equality between the sum of the $n_k$ and the sum of the $\mu_k(m_\chi)$ for
any two kinematic distributions being compared.  Thus, the $G(m_\chi)$ statistic
for each $m_\chi$ in the survey takes the form given in 
Eq.~(\ref{eq:DefOfGStatistic}), rather than the alternative form appropriate for data
distributed according to a multinomial distribution.

\begin{table*}
\begin{center}
\begin{tabular}{|cc|cccc|}\hline
\multicolumn{2}{|c|}{~~~DDM Benchmark~~~} & 
\multicolumn{4}{|c|}{Significance $\sigma$} \\ \cline{1-6} 
~$\Delta m$~ & ~$\gamma$~ & 
~$H_{T_{jj}}^{\mathrm{min}} = 275$~GeV~ & 
~$H_{T_{jj}}^{\mathrm{min}} = 325$~GeV~ &
~$H_{T_{jj}}^{\mathrm{min}} = 375$~GeV~ & 
~$H_{T_{jj}}^{\mathrm{min}} = 425$~GeV~ \\          
\hline
~50~GeV~  & 0 & 0.03 & 0.02 & 0.01 & 0.01 \\
~50~GeV~  & 1 & 3.08 & 2.60 & 1.20 & 0.22 \\
~50~GeV~  & 2 & 3.13 & 1.35 & 0.09 & 0.00 \\ \hline
~300~GeV~ & 0 & 0.00 & 0.00 & 0.00 & 0.00 \\
~300~GeV~ & 1 & 1.39 & 1.33 & 1.22 & 1.17 \\
~300~GeV~ & 2 & 4.63 & 4.14 & 3.02 & 1.31 \\ \hline
~500~GeV~ & 0 & 0.02 & 0.01 & 0.01 & 0.01 \\
~500~GeV~ & 1 & 0.00 & 0.00 & 0.00 & 0.00 \\
~500~GeV~ & 2 & 0.00 & 0.00 & 0.00 & 0.00 \\
\hline
\end{tabular}
\caption{ ~The statistical significance of differentiation derived from examining the 
goodness of fit between the $\MT2$ distributions associated with a variety of benchmark 
DDM models and those associated with traditional dark-matter models.  The
values of $\Delta m$ and $\gamma$ for each DDM model are specified in the table, and in all
cases we have taken $m_0 = 100$~GeV and $\delta = 1$.  The results shown here correspond 
to an integrated luminosity of 
$\mathcal{L}_{\mathrm{int}} = 300\mathrm{~fb}^{-1}$ in each of the two LHC detectors
for a center-of-mass energy $\sqrt{s} = 14$~TeV.  The event-selection criteria imposed 
include the minimum cut on $H_{T_{jj}}$ shown in the table as well as the other
selection cuts discussed in the text.  
\label{tab:FinalSigsForHTjjCut}}
\end{center}
\end{table*}

Having defined the $G_{\mathrm{min}}$ statistic in Eq.~(\ref{eq:DefOfGmax}),
we now turn to examine how the imposition of event-selection criteria can affect the 
distinctiveness of the distributions of key kinematic variables ---
in particular, $\MT2$ and $\met$.  We begin by examining the effect 
of imposing a minimum cut $\alpha_T^{\mathrm{min}}$ on $\alpha_T$.   
In the top left panel of Fig.~\ref{fig:GminCurves}, we plot the value of $G_{\mathrm{min}}$
for the $\MT2$ distributions associated with several DDM ensembles as a function of
$\alpha_T^{\mathrm{min}}$.  The precuts are the only additional cuts imposed on the data.
The results shown here correspond to an integrated luminosity
$\mathcal{L}_{\mathrm{int}} = 300\mathrm{~fb}^{-1}$ at each of the LHC detectors.   
Note that while $G_{\mathrm{min}}$ generally decreases with 
increasing $\alpha_T^{\mathrm{min}}$ due to the overall reduction in number of 
signal events, it does not do so monotonically.  Indeed, for all curves shown, 
$G_{\mathrm{min}}$ actually {\it rises}\/ with increasing $\alpha_T^{\mathrm{min}}$ within
the range $0.4 \lesssim \alpha_T^{\mathrm{min}} \lesssim 0.55$.  As can readily be 
seen from Fig.~\ref{fig:ScatterPlotsMT2}, this is precisely the range within which
the $\alpha_T$ cut effectively eliminates the region of parameter space within which 
the data-point distributions corresponding to different invisible-particle masses
overlap, yet still retains the majority of the events in the region
within which those data-point distributions are the most distinctive.  A similar 
enhancement in the $G_{\mathrm{min}}$ values obtained from the corresponding $\met$ 
distributions is apparent in the bottom left panel of Fig.~\ref{fig:GminCurves}.   

We now turn to examine the effect on $G_{\mathrm{min}}$ of imposing a minimum
cut $H_{T_{jj}}^{\mathrm{min}}$ on $H_{T_{jj}}$.  In the top right panel of 
Fig.~\ref{fig:GminCurves}, we plot the value of $G_{\mathrm{min}}$ for the $\MT2$ 
distributions associated with the same DDM ensembles as a function of 
$H_{T_{jj}}^{\mathrm{min}}$.  Once again, the precuts are the only additional cuts 
imposed on the data.  In contrast with the $G_{\mathrm{min}}$ curves for 
$\alpha_T^{\mathrm{min}}$, we see that the corresponding curves for 
$H_{T_{jj}}^{\mathrm{min}}$ fall monotonically due to the loss in statistics.  This 
is to be expected, as we have seen that there is no advantageous correlation between 
$H_{T_{jj}}$ and $\MT2$ which can be exploited to offset this loss.  The same behavior 
is also apparent in the $G_{\mathrm{min}}$ values obtained from the corresponding $\met$ 
distributions in the bottom right panel of Fig.~\ref{fig:GminCurves}.

Despite the enhancement within the range 
$0.4 \lesssim \alpha_T^{\mathrm{min}} \lesssim 0.55$ discussed above, it is 
nevertheless clear from Fig.~\ref{fig:GminCurves} 
that increasing $\alpha_T^{\mathrm{min}}$ generally has the effect of diminishing 
our power to discriminate non-minimal dark sectors from traditional dark sectors.
Likewise, we see that a similar 
conclusion holds, perhaps even more dramatically, for cuts on $H_{T_{jj}}$.  
However, the extraction of signal from background typically requires more 
than simply one or the other cut in isolation:  we typically need to impose an 
entire slew of cuts simultaneously.  Inevitably, these cuts, which are 
designed to enhance signal extraction, further reduce our power to resolve 
non-trivial dark sectors relative to traditional dark sectors.  Thus, we find 
ourselves in a position in which we must ultimately balance considerations 
related to signal extraction against those related to dark-sector resolution.

In order to study this issue, we adopt a set of cuts which is similar to 
those employed in the CMS jets~+~$\met$ analysis in Ref.~\cite{CMSJetsPlusMET}.  
In particular, we simultaneously require that  
\begin{itemize}
  \item $\{p_{T_{j_1}},p_{T_{j_2}}\} \geq 100$~GeV
  \item $\met \geq 90$~GeV
  \item $\alpha_T \geq 0.55$~.  
\end{itemize}
In addition to these cuts, we also impose a minimum cut of the form 
$H_{T_{jj}} \geq H_{T_{jj}}^{\mathrm{min}}$ on the data and examine the effect of
varying $H_{T_{jj}}^{\mathrm{min}}$ on the statistical significance of differentiation
between DDM and traditional dark-matter models.  As discussed in Sect.~\ref{sec:CorrelBetweenVars}, a 
cut of this sort can be effective in reducing the backgrounds 
from SM processes such as $t\bar{t}$~+~jets, $W^\pm$~+~jets, and $Z$~+~jets ---
processes which can give rise to genuine sources of missing energy in the form 
of neutrinos, and whose contributions to the total SM background are therefore more
likely to survive the $\alpha_T$ cut.  Note also that these selection cuts are
sufficient for passing CMS triggering requirements, provided that 
$H_{T_{jj}}^{\mathrm{min}} \geq 275$~GeV~\cite{CMSJetsPlusMET}.

Our results are shown in Table~\ref{tab:FinalSigsForHTjjCut}, 
where we display the Gaussian-equivalent 
significance of differentiation for the $\MT2$ distributions associated with 
several benchmark DDM models after the imposition of the selection cuts described 
above.  These benchmark models are characterized by different values of $\Delta m$ 
and $\gamma$, with fixed $m_0 = 100$~GeV.  Results are given for several 
different choices of $H_{T_{jj}}^{\mathrm{min}}$.

The decline in sensitivity with increased $H_{T_{jj}}^{\mathrm{min}}$ can be 
qualitatively understood as follows.
Since events involving heavier $\chi_n$ tend to have both smaller $H_{T_{jj}}$ values
and smaller $\MT2$ values, as shown in the bottom left panel of 
Fig.~\ref{fig:ScatterPlotsMT2}, increasing $H_{T_{jj}}^{\mathrm{min}}$ results in a 
disproportionately severe reduction in events involving heavier $\chi_n$ in comparison 
with events involving lighter ones.  As a result, increasing $H_{T_{jj}}^{\mathrm{min}}$ 
has the effect of washing out the imprints of the heavier $\chi_n$ in $\MT2$ distributions 
and leads to a decrease in the significance of differentiation, as seen in 
Table~\ref{tab:FinalSigsForHTjjCut}.  

Similarly, the dependence of the results shown in Table~\ref{tab:FinalSigsForHTjjCut} 
on the power-law index $\gamma$ can be qualitatively understood as follows.
As $\gamma$ decreases, the width $\Gamma_\phi$ of $\phi$ will be increasingly 
dominated by the contribution from decays to the lightest dark-matter component $\chi_0$. 
Thus, for sufficiently small $\gamma$, the resulting kinematic distributions 
become effectively indistinguishable from those obtained for a traditional dark-matter 
candidate of mass $m_\chi = m_0$.  Conversely, for sufficiently large $\gamma$, it turns 
out that $\Gamma_\phi$ will be dominated by the contributions from the most massive
kinematically accessible states in the ensemble.  In this regime,
the resulting kinematic distributions become effectively indistinguishable from 
those obtained for a traditional dark-matter candidate with $m_\chi$ equal to the 
mass of the most massive accessible ensemble component.  Indeed, as was noted in 
Ref.~\cite{DDMLHC1} in the case of three-body parent-particle decays, there 
exists a particular intermediate range of $\gamma$ for any particular assignment of 
$m_0$, $\Delta m$, \etc, within which the branching fractions of $\phi$ to two or 
more of the $\chi_n$ are of roughly the same order and the corresponding 
kinematic distributions are therefore more distinctive. 

Taken together, then, the primary message of the results shown in 
Table~\ref{tab:FinalSigsForHTjjCut} is that there are 
non-trivial regions of our parameter space within which the population of signal events 
associated with a DDM ensemble can be distinguished from the population of signal events 
associated with any traditional dark-matter candidate on the basis of the distributions 
of kinematic variables such as $\MT2$.  Indeed, with 
an integrated luminosity of $300\mathrm{~fb}^{-1}$ in both LHC detectors, a statistical 
significance of differentiation close to $5\sigma$ is obtained for low-to-moderate 
values of $H_{T_{jj}}^{\mathrm{min}}$ for 
$50\mathrm{~GeV} \lesssim m_0 \lesssim 300\mathrm{~GeV}$ and 
$1 \lesssim \gamma \lesssim 2$.


\section{Discussion and Conclusions\label{sec:conclusions}}


In this paper, we have investigated the prospects for distinguishing non-minimal 
dark sectors in the dijet~+~$\met$ channel at the LHC.~  Almost by necessity, 
searches of this sort --- both in this channel and others --- involve not merely 
identifying an excess in the total number of signal events over background, but actually 
analyzing the shapes of the full distributions of the relevant kinematic variables.
It is therefore critical to examine the correlations between such variables, 
since cuts imposed on one variable in order to reduce the background have the capacity 
to alter or distort the distributions of other variables which are critical for probing 
the structure of the dark sector.      
 
Using DDM ensembles as a benchmark, we have examined the extent to which the 
distributions of different kinematic variables are impacted by dark-sector 
non-minimality.  We have shown that the distributions of certain 
variables such as $\met$ and $\MT2$ are particularly sensitive to the properties 
of the dark-sector particles.  By contrast, we have shown that the distributions of 
other variables such as $\alpha_T$ and $|\Delta\phi_{jj}|$ are comparatively 
insensitive to the details of the dark sector.  Finally, we have shown that still 
other variables such as $H_{T_{jj}}$ lie between these extremes. 
 
Furthermore, we have also demonstrated that non-trivial correlations exist between the 
variables $\lbrace \met,\MT2 \rbrace$ and the variables 
$\lbrace \alpha_T,|\Delta\phi_{jj}|, H_{T_{jj}}\rbrace$.  In particular, we find 
that $\alpha_T$ is correlated with $\MT2$ and $\met$ in such a way that a threshold 
cut on $\alpha_T$ can actually {\it enhance}\/ the distinctiveness of 
the corresponding kinematic distributions in certain situations.  Indeed, we
have shown that this effect more than offsets the corresponding loss in statistics
for certain values of $\alpha_T^{\mathrm{min}}$.  By contrast, 
we find that $H_{T_{jj}}$ is correlated with these same variables in such a way 
that a threshold cut on $H_{T_{jj}}$ generically serves to wash out distinguishing 
features in the corresponding distributions.  

Finally, we have investigated the impact of such cuts of the distinctiveness
of the $\met$ and $\MT2$ distributions associated with our DDM ensembles, as 
quantified by the goodness-of-fit statistic $G_{\mathrm{min}}$.  
We have shown that correlations between variables give rise to a non-trivial 
dependence of $G_{\mathrm{min}}$ on the cuts imposed --- a dependence which
transcends mere issues of event count.  Due in part to these effects,
the signal-event distributions associated with DDM ensembles can be   
distinguished from those associated with traditional dark-matter candidates 
at a significance level approaching $5\sigma$ in many situations. 
   
One final comment is in order.
Our focus in this paper has been on the correlations between selection cuts and 
kinematic distributions of signal events and on the effects that such correlations have on 
the distinctiveness of those distributions.  We have therefore focused our analysis 
primarily on the signal contributions from different dark-sector models alone and have only 
incorporated the SM backgrounds into our analysis as a motivation for the cuts imposed
on certain kinematic variables.  However, despite the established efficiency of the selection cuts 
adopted here in reducing those backgrounds~\cite{CMSJetsPlusMET,CMSJetsPlusMETUpdate} (and 
especially the contribution from QCD processes), we note that the residual backgrounds from 
$\bar{t}t$~+~jets, $W^\pm$~+~jets, and $Z$~+~jets are still quite sizable.  Nevertheless, it 
may be possible to isolate these residual backgrounds using other techniques.  
For example, both the normalization and shape of the ``irreducible'' background from 
$Z +$~jets can in principle be determined from the related process in which the $Z$ decays 
into a pair of charged leptons~\cite{RandallDijetVariables}.  Such information could in principle 
allow for a modeling of this background that would make background subtraction a viable
possibility.  Further reducing the $W^\pm$~+~jets background is a
significantly more challenging endeavor.  However, while a full analysis of the effect
of selection cuts on the combined contribution from both signal and background processes to 
the relevant kinematic distributions is beyond the scope of this paper, we emphasize 
that the correlations we have investigated here are every bit as relevant for such a study
as they have been for this background-free analysis.


\begin{acknowledgments}


We would like to thank P.~Loch for useful discussions.  KRD and SS are supported in 
part under DOE Grant DE-FG02-13ER-41976, while BT is supported in part by 
NSERC Canada.  SS also wishes to acknowledge the hospitality of the Aspen Center 
for Physics, which is supported in part under NSF Grant PHY-1066293.  
The opinions and conclusions expressed herein are those of the authors, 
and do not represent any funding agency.
\end{acknowledgments}


\bigskip


\end{document}